\documentclass[aps,prb,twocolumn,groupedaddress,floatfix,amsmath,amssymb,superscriptaddress]{revtex4-2}
\usepackage{amsmath}
\usepackage{amssymb}
\usepackage{braket}
\usepackage{graphicx}
\usepackage[table]{xcolor}
\usepackage{verbatim}
\usepackage{float}
\usepackage{color}
\usepackage{siunitx}
\usepackage{gensymb}
\usepackage{bm}
\usepackage{xcolor}
\usepackage{multirow}
\usepackage{footnote}
\usepackage{bbm}

\usepackage{array}
\newcolumntype{P}[1]{>{\centering\arraybackslash}p{#1}}
\newcolumntype{Q}[1]{>{\raggedleft\arraybackslash}p{#1}}
\newcolumntype{R}[1]{>{\raggedright\arraybackslash}p{#1}}

\newcommand{\rr}{\mathbf{r}}

\newcommand{\qq}{\mathbf{q}}
\newcommand{\QQ}{\mathbf{Q}}
\newcommand{\kk}{\mathbf{k}}

\newcommand{\gG}{\mathbf{g}}
\newcommand{\bvec}[1]{\mathbf{\boldsymbol{#1}}}

\newcommand{\rrho}{\boldsymbol{\rho}}
\newcommand{\unitx}{\hat{\mathbf{x}}}
\newcommand{\unity}{\hat{\mathbf{y}}}
\newcommand{\Scal}{\mathcal{S}}
\newcommand{\Xcal}{\mathcal{X}}
\newcommand{\Ecal}{\mathcal{E}}
\newcommand{\Hcal}{\mathcal{H}}

\newcommand{\braoket}[3]
{
	\big< #1 \big| #2 \big| #3 \big>
}

\begin{document}
\title{Dimensionality crossover for moir\'e excitons in twisted bilayers of anisotropic two-dimensional semiconductors}

\author{Isaac Soltero}
\affiliation{Departamento de F\'isica Qu\'imica, Instituto de F\'isica, Universidad Nacional Aut\'onoma de M\'exico, Ciudad de M\'exico, C.P. 04510, M\'exico}
\affiliation{Department of Physics and Astronomy, University of Manchester. Booth St.\ E., Manchester, M13 9PL, United Kingdom}
\affiliation{National Graphene Institute, University of Manchester. Booth St.\ E., Manchester, M13 9PL, United Kingdom}
\author{David A.\ Ruiz-Tijerina}
\email{d.ruiz-tijerina@fisica.unam.mx}
\affiliation{Departamento de F\'isica Qu\'imica, Instituto de F\'isica, Universidad Nacional Aut\'onoma de M\'exico, Ciudad de M\'exico, C.P. 04510, M\'exico}

\begin{abstract}
We study the energies and optical spectra of excitons in twisted bilayers of anisotropic van der Waals semiconductors exhibiting moir\'e patterns, taking phosphorene as a case study. Following the electronic Hamiltonian introduced by us in Ref.\ \onlinecite{MoirePhos.PhysRevB.105.235421}, and leveraging the scale separation between the moir\'e lengthscale and the exciton Bohr radii, we introduce a continuous model for excitons that incorporates the spatial variation of their binding energies. Our zone-folding calculations reveal a dimensionality crossover for the exciton states, driven by the combined dispersion- and moir\'e potential anisotropies, from quantum-dot-like (0D) lattices at twist angles $\theta<\theta_*$, to quantum-wire-like (1D) arrays at $\theta>\theta_*$, with crossover angle $\theta_*=4^\circ$. We identify clear signatures of this dimensionality crossover in the twist angle dependence of the excitonic absorption spectra, which allows experimental verification of our theoretical results through standard optical measurements. Our results establish two-dimensional anisotropic moir\'e semiconductors as versatile solid-state platforms for exploring bosonic correlations across different dimensionalities.
\end{abstract}

\maketitle

\textit{Introduction.}---Moir\'e heterostructures of two-dimensional (2D) semiconductors have recently emerged as solid-state quantum simulators\cite{CMquantumsim2021}, exhibiting multiple strongly-correlated states of fermionic\cite{wang2020correlated,tang2020simulation,li2021imaging,wang2022one}, bosonic\cite{ma2021strongly,gu2022dipolar}, and mixed Fermi-Bose\cite{zeng2023exciton} matter. Heterostructures based on hexagonal crystals, such as transition-metal dichalcogenides (TMDs), are known to realize generalized Hubbard models for charge carriers and excitons\cite{wu2018hubbard,angeli2021gamma,multifaceted}, where strong correlations arise due to large on-site interaction to tunnelling ratios between neighboring superlattice sites, controlled by the twist angle. More recently, anisotropic moir\'e semiconductors with rectangular unit cells have been predicted\cite{RubioGeSe,KariyadoGeSe,MoirePhos.PhysRevB.105.235421} and experimentally verified\cite{wang2022one} to host electronic Tomonaga-Luttinger liquids. In these materials, including phosphorene, group-IV monochalcogenides (\emph{e.g.}~GeSe and SnSe), and $1T'$-phase TMDs, the moir\'e superlattice strongly amplifies the structural and band anisotropies, resulting in the formation of one-dimensional (1D) conduction and valence states with quantum-wire-like spatial profiles, where strong correlations arise due to the strong lateral confinement\cite{giamarchi2003quantum}. Crossovers from this Tomonaga-Luttinger regime into both an anisotropic Hubbard- and a 2D dispersive regime are possible by decreasing and increasing the interlayer twist angle, respectively, as predicted in Refs.\ \onlinecite{MoirePhos.PhysRevB.105.235421,guo2023pseudo}, making moir\'e anisotropic semiconductors promising platforms for exploring strong correlations between zero-, one- and two-dimensional fermions. However, the effects of these dimensionality crossovers remains unexplored in the case of the ubiquitous bosonic quasiparticle arising in 2D semiconductors: the exciton.

In this Letter, twisted bilayers of anisotropic 2D semiconductors are established as versatile platforms for exploring excitonic physics across dimensionalities. We introduce a fully parametrized continuous Hamiltonian for excitons in the resulting moir\'e superlattice, based on our model for carriers of Ref.\ \onlinecite{MoirePhos.PhysRevB.105.235421}, and which relies only on a clear scale separation between the excitonic Bohr radii and the moir\'e wavelength. Numerical solution of our model using zone-folding methods reveals a dimensionality crossover for the low-energy excitons, going from 0D, quantum-dot-like states at small twist angles $\theta < \theta_*$, to 1D quantum-wire-like states for $\theta > \theta_*$, with a theoretical crossover angle $\theta_*=4^\circ$. The twist-angle-dependent excitonic absorption spectrum, computed here within a linear response approximation, bears clear signatures of the predicted dimensionality crossover, in the form of a sharp slope change for the first absorption peak---which blueshifts linearly with increasing twist angle---precisely at the crossover value $\theta_*$. Whereas the symmetry-based model discussed in this Letter was specifically parametrized for twisted phosphorene bilayers, we expect that our results can be extended to excitons in moir\'e heterostructures formed with monolayers of group-IV monochalcogenides, such as GeSe and SnSe, which differ structurally from phosphorene only in their lack of inversion symmetry. Whereas lateral confinement of 2D excitons has been previously observed\cite{bai2020excitons} in strained TMD systems,  twisted bilayers of phosphorene, and potentially of group-IV monochalcogenides, offer a twist-angle controled crossover between 0D and 1D exciton states in their equilibrium state. Our results establish twisted bilayers of anisotropic 2D semiconductors as strong solid-state candidates for quantum simulators of interacting bosons across dimensionalities\cite{coldatoms1,BoseLuttingerVishwanath,giamarchi1d2d}.

\begin{figure}
    \centering
    \includegraphics[width=1.0\columnwidth]{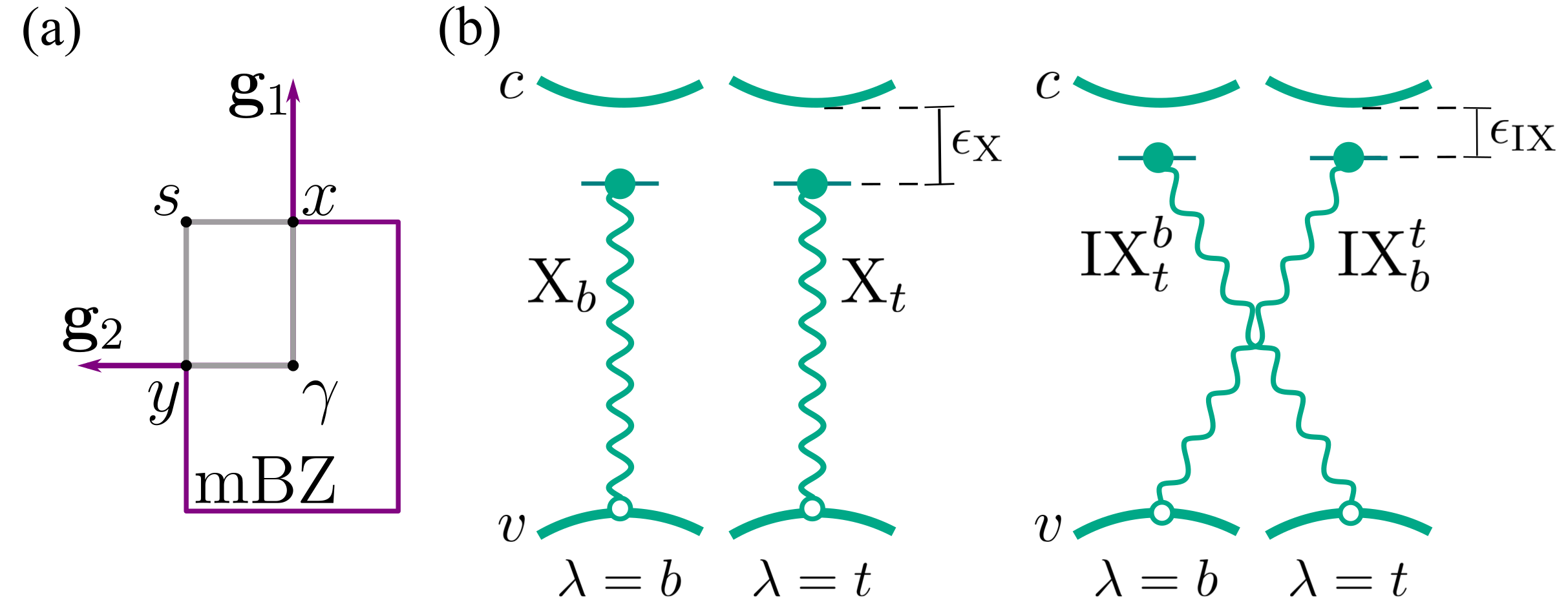}
    \caption{(a) Moiré Bragg vectors and the corresponding moiré Brillouin zone (mBZ). Symmetry points are labeled as $\gamma$, $x$, $s$ and $y$. (b) Schematic of the two possible intra- and interlayer excitons. Electrons (holes) are indicated by a solid (empty) circle. Wavy lines indicate the electrostatic interaction binding the electron-hole pair, with a binding energy $\epsilon_{\rm X}$ or $\epsilon_{\rm IX}$.}
    \label{fig:ExcConfig}
\end{figure}

\textit{Model.}---We consider the moiré pattern formed in a phosphorene bilayer with a small relative twist angle $\theta \lesssim 6^\circ$, corresponding to a large moir\'e supercell (mSC), containing over $100$ atomic unit cells. Every region in the mSC, centered at some position $\rr$ along the sample plane, is locally described by its approximate commensurate stacking, fully defined by an in-plane offset vector $\rr_{0}(\rr)$, and the local interlayer distance $d[\rr_{0}(\rr)]$. The reciprocal-space primitive vectors of the moiré superlattice (mSL) can be approximated as
\begin{equation}
    \gG_{1}\approx \frac{2\pi\theta}{a_{x}}\,\hat{\bvec{\mathrm{y}}},\quad \gG_{2}\approx -\frac{2\pi\theta}{a_{y}}\,\hat{\bvec{\mathrm{x}}},
\end{equation}
and define the moir\'e Brillouin zone (mBZ) shown in Fig.\ \ref{fig:ExcConfig}(a), with $a_{x}=3.296$ $\text{\AA}$ and $a_{y}=4.590$ $\text{\AA}$ the monolayer lattice constants extracted from \textit{ab initio} calculations \cite{MoirePhos.PhysRevB.105.235421}. The corresponding mSL basis vectors are
\begin{equation}
    \bvec{a}_{1}^{\rm M} \approx \frac{a_{x}}{\theta}\,\hat{\bvec{\mathrm{y}}}, \quad \bvec{a}_{2}^{\rm M} \approx -\frac{a_{y}}{\theta}\,\hat{\bvec{\mathrm{x}}}.
\end{equation}

To study the exciton states of the twisted bilayer, we start from the low-energy continuous model of the moir\'e potential for $\Gamma$-point conduction- ($c$) and valence ($v$) electrons of both layers, introduced in Ref.\ \onlinecite{MoirePhos.PhysRevB.105.235421}:
\begin{equation}\label{ElecHamReal}
\begin{split}
    H_{\rm m} =&\, \sum_{\alpha,\lambda} \int d^{2}r\, \varepsilon_{\alpha}^{\lambda}(\rr)\varphi_{\alpha\lambda}^{\dagger}(\rr)\varphi_{\alpha\lambda}(\rr) \\
    & + \sum_{\alpha} \int d^{2}r\, \big[ T_{\alpha}(\rr)\varphi_{\alpha t}^{\dagger}(\rr)\varphi_{\alpha b}(\rr) + {\rm H.c.}\big], \\
\end{split}
\end{equation}
with $\varphi_{\alpha\lambda}(\rr)$ the electron field operator for band $\alpha=c,v$ in layer $\lambda=t,\,b$ (for top and bottom, respectively) at position $\rr$. The position-dependent state energies $\varepsilon_{\alpha}^{\lambda}(\rr)$ and tunneling energies $T_{\alpha}(\rr)$ have the mSL periodicity, and as such are expressed as Fourier series over the mSL reciprocal vectors $\gG_{m,n}=m\gG_1+n\gG_2$, with $m,\,n$ integers (see Supplementary Material).

We then evaluate the matrix elements of the moir\'e potential \eqref{ElecHamReal} between the different exciton states of interest that can be formed in the four-band system. We identify two types of intralayer excitons (X)---one for each monolayer---, and two types of interlayer excitons (IX), shown schematically in Fig.\ \ref{fig:ExcConfig}(b), with two-body wave functions
\begin{subequations}\label{ExcStates}
\begin{equation}\label{eq:Xbasis}
\begin{split}
    \ket{\text{X}_{\lambda,n}(\QQ)} = \int & d^2r\,\frac{e^{i\QQ\cdot\rr}}{\sqrt{\Scal}}\int d^2\rho\,X_n(\rho)\\ &\times\varphi_{c\lambda}^\dagger(\rr_e[\rrho,\rr])\varphi_{v\lambda}(\rr_h[\rrho,\rr])\ket{\Omega},
\end{split}
\end{equation}
\begin{equation}\label{eq:IXbasis}
\begin{split}
    \ket{\text{IX}_{\lambda,n}^{\bar{\lambda}}(\QQ)} = \int & d^2r\,\frac{e^{i\QQ\cdot\rr}}{\sqrt{\Scal}}\int d^2\rho\,Y_n(\rho)\\ &\times\varphi_{c\bar{\lambda}}^\dagger(\rr_e[\rrho,\rr])\varphi_{v\lambda}(\rr_h[\rrho,\rr])\ket{\Omega}.
\end{split}
\end{equation}
\end{subequations}
Here, $\ket{\text{X}_{\lambda,n}(\QQ)}$ represents an intralayer exciton with relative motion (RM) quantum numbers $n$, and center of mass (CoM) wave vector $\QQ$ in layer $\lambda$; and $\ket{\text{IX}_{\lambda,n}^{\bar{\lambda}} (\QQ)}$ an interlayer exciton formed by a $\lambda$-layer hole, and an electron in the opposite layer $\bar{\lambda}$. $\rr$ and $\rrho$ are the CoM and RM position vectors, respectively, which determine the electron and hole positions $\rr_{e}$ and $\rr_{h}$ (see Supplementary Material); $X_n(\rrho)$ and $Y_n(\rrho)$ are the corresponding electron-hole RM wave functions; and $\mathcal{S}$ is the sample surface area.

The exciton binding energies and RM wave functions are described by the anisotropic Wannier equation, with a screened electrostatic interaction corresponding to a bilayer immersed in a medium with dielectric tensor $\epsilon = \text{diag}(\epsilon_{\parallel}, \epsilon_{\parallel}, \epsilon_{\perp})$. Given its experimental relevance, we will consider hexagonal boron nitride encapsulation (hBN, $\epsilon_{\parallel} = 6.9$, $\epsilon_{\perp} = 3.7$ \cite{geick1966normal,laturia2018dielectric}) for the phosphorene bilayer. The electron-hole interactions in the bilayer depend on the interlayer distance\cite{danovich2018localized,viner2021excited} $d[\rr_0(\rr)]$, which varies spatially according to the local stacking $\rr_0(\rr)$ (see Supplementary Material), thus making the binding energies and RM wave functions position dependent within the continuous approximation. As the stacking vector $\rr_0(\rr)$ varies slowly across the moir\'e supercell, over length scales of the order of the moir\'e periodicity, so does the interlayer distance. By comparison, the excitonic RM wavefunction extension is only $\sim 10\,{\rm \AA}$ \cite{henriques2020excitons}. This clear scale separation allows us to treat the exciton binding energies as adiabatic functions of position, effectively representing scalar potentials for intra- and interlayer excitons\footnote{We have also considered the RM wave functions as adiabatically depending on the local stacking configuration: $X[\rrho,\rr_0(\rr)]$ and $Y[\rrho,\rr_0(\rr)]$. The $\rr_0$-dependence of these functions introduces an additional spatial dependence to the matrix elements $\braoket{{\rm X}_{\lambda',n'}(\QQ')}{H_{\rm exc}}{{\rm IX}_{\lambda,n}^{\bar{\lambda}}(\QQ)}$.  We have numerically determined that this variation is $<1\%$, and thus negligible.}, $\epsilon_{\rm X}(\rr)$ and $\epsilon_{\rm IX}(\rr)$, respectively.

\begin{figure*}[t!]
    \begin{center}
    \includegraphics[width=1.0\textwidth]{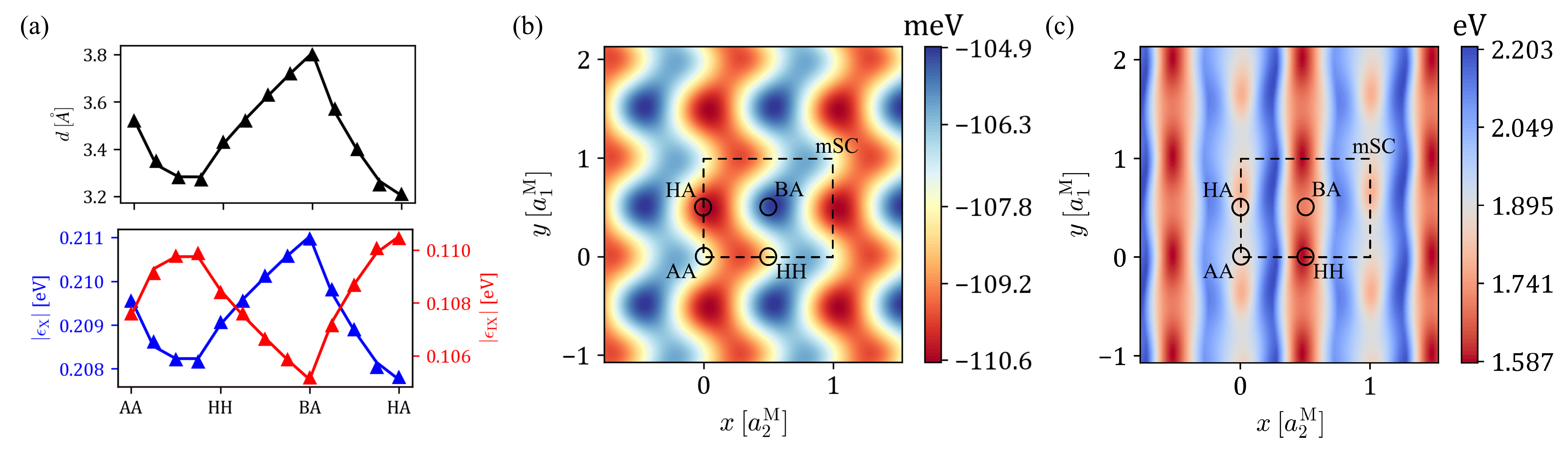}
    \caption{(a) Interlayer distance $d$ (top) and absolute value of exciton binding energies $|\varepsilon|$ (bottom) at 13 different stacking configurations across the mSC, including the high-symmetry points AA, HH, BA and HA. For both intralayer ($\varepsilon_{\rm X}$, blue) and interlayer ($\varepsilon_{\rm IX}$, red) binding energies the fitting function \eqref{FourEb} is shown as solid lines with parameters shown in Table \ref{tab:EFitParam}. (b) Interlayer exciton binding energy as a function of position for the twisted phosphorene bilayer resulting from the interpolation \eqref{FourEb}. (c) Effective moiré potential for excitons, where potential wells are formed at the HH mSL regions. For both (b) and (c), $x$ and $y$ axes are scaled with respect to the superlattice parameters $a_{2}^{\rm M}$ and $a_{1}^{\rm M}$, respectively. The mSC and high-symmetry regions are labeled.}
    \label{fig:Panel1}
    \end{center}
\end{figure*}

The moiré pattern contains four distinct stacking configurations with a particular symmetry point group (high symmetry regions), which we label AA, HH, HA and BA in Fig.\ \ref{fig:Panel1}. We have solved the anisotropic Wannier equation locally at these four, and nine other intermediate regions of the mSC, for a total of thirteen representative stackings, using a semi-analytical direct diagonalization method \cite{griffin1957collective} that has proven successful for studying excitons in 2D semiconductors \cite{henriques2019optical,henriques2020excitons,ruiz2020theory,viner2021excited} (see Supplementary material). The calculated local binding energies across the mSC are reported in Fig.\ \ref{fig:Panel1}(a), along with the corresponding interlayer distances, from Ref.\ \onlinecite{MoirePhos.PhysRevB.105.235421}. Whereas a Rydberg-like sequence of exciton states is obtained at each local stacking configuration, here we focus only on the lowest X and IX states, which we henceforth call $1s$ excitons\footnote{The lowest X and IX states transform as the $A_1$ irreducible representation of the symmetry group $C_{2v}$ of the RM Hamiltonian, and look like $1s$ hydrogenic states elongated in the $y$ direction (see Supplementary Material), justifying the label $1s$. The eventual importance of, \emph{e.g.}, $2p$ or $2s$ excited states for the $1s$ moir\'e exciton band structures is determined by the $2p-2s$ and $2s-1s$ wave function overlaps, which we estimate to be at least one order of magnitude smaller than any $1s-1s$ overlap. Moreover, the oscillator strength of the $2s$ intralayer exciton is also estimated to be much weaker than that of its $1s$ counterpart, such that it can be neglected in the optical spectrum.}. Figure \ref{fig:Panel1}(a) shows opposite trends for the X and IX binding energies as functions of the interlayer distance, which can be understood as follows: the screening by layer $\bar{\lambda}$ of the electron-hole interaction in layer $\lambda$ is reduced as $d$ increases, leading to a larger $|\epsilon_{\rm X}|$. By contrast, a larger $d$ increases the electron-hole separation in an interlayer exciton, in detriment of the interlayer interaction, thus reducing $|\epsilon_{\rm IX}|$. The scalar potentials $\epsilon_\mu(\rr)$ ($\mu={\rm X,\,IX}$) are obtained by interpolating the stacking dependence of the binding energies through the formula
\begin{equation}\label{FourEb}
    \epsilon_\mu(\rr) = \epsilon_{\mu,0} + \sum_{n=1}^{N}\big[ \epsilon_{\mu,n}^{s}\cos(\gG_{n}\cdot\rr) + \epsilon_{\mu,n}^{a}\sin(\gG_{n}\cdot\rr)\big].
\end{equation}
Good agreement between Eq.\ \eqref{FourEb} and the numerical results is obtained for $N=4$, with the fitting parameters of Table \ref{tab:EFitParam}, as shown with solid lines in Fig.\ \ref{fig:Panel1}(a). The spatial variation of the extrapolated IX binding energy \eqref{FourEb} across the moir\'e superlattice is shown in Fig.\ \ref{fig:Panel1}(b).

\begin{table}[b!]
    \caption{Binding energy interpolation parameters in Eq.\ \eqref{FourEb} for intra- (X) and interlayer (IX) excitons. All parameters are reported in meV.}
    \centering
    \begin{tabular}{P{1.1cm} | Q{0.7cm}@{.}R{0.9cm} Q{0.7cm}@{.}R{0.9cm} | Q{0.7cm}@{.}R{0.9cm} Q{0.7cm}@{.}R{0.9cm}}
    \hline
    \hline
        \, & \multicolumn{4}{c}{X} & \multicolumn{4}{c}{IX}\\
        \raisebox{1.5ex}[0pt]{$n$}  & \multicolumn{2}{c}{$\epsilon_{{\rm X},n}^{s}$} & \multicolumn{2}{c}{$\epsilon_{{\rm X},n}^{a}$} & \multicolumn{2}{c}{$\epsilon_{{\rm IX},n}^{s}$} & \multicolumn{2}{c}{$\epsilon_{{\rm IX},n}^{a}$} \\
        \hline
        1 & 0&030 & $-$0&157 & $-$0&072 & $-$0&142 \\
        2 & 0&714 & $-$0&800 & $-$1&189 & 1&336 \\
        3 & $-$0&466 & 0&047 & 0&757 & $-$0&268 \\
        4 & $-$0&477 & $-$0&296 & 0&809 & 0&267 \\
        \hline
         & \multicolumn{4}{c}{$\epsilon_{{\rm X},0} = -209.392$} & \multicolumn{4}{c}{$\epsilon_{{\rm IX},0} = -107.833$} \\
    \hline
    \hline
    \end{tabular}
    \label{tab:EFitParam}
\end{table}

Computing the matrix elements of \eqref{ElecHamReal}, including  \eqref{FourEb}, in the two-particle basis \eqref{ExcStates}, we arrive at the following representation for the effective moir\'e potential for excitons:
\begin{equation}\label{ExcHam}
    \Hcal_{m}(\rr) =
    \begin{pmatrix}
    \mathcal{E}_{\text{X}}(\rr) & 0 & \tilde{T}_{c}(\rr) & -\tilde{T}_{v}(\rr) \\
    0 & \mathcal{E}_{\text{X}}(\rr) & -\tilde{T}_{v}(\rr) & \tilde{T}_{c}(\rr) \\
    \tilde{T}_{c}(\rr) & -\tilde{T}_{v}(\rr) & \mathcal{E}_{\text{IX}}(\rr)  & 0 \\
    -\tilde{T}_{v}(\rr) & \tilde{T}_{c}(\rr) & 0 & \mathcal{E}_{\text{IX}}(\rr)  
    \end{pmatrix},
\end{equation}
with the basis ordering $\{ \ket{\text{X}_{b}},\ket{\text{X}_{t}},\ket{\text{IX}_{b}^{t}},\ket{\text{IX}_{t}^{b}}\}$, and with tunneling functions $\tilde{T}_{\alpha}$, renormalized with respect to their single-particle counterparts by the numerically computed overlap between the intra- and interlayer RM wave functions. We have defined the potentials
\begin{equation}
    \mathcal{E}_{\mu}(\rr) = \mathcal{E}^{(0)} + \delta\varepsilon_{c}(\rr) - \delta\varepsilon_{v}(\rr) + \epsilon_{\mu}(\rr),
\end{equation}
containing the position-dependent conduction- and valence band edge energies $\delta\varepsilon_\alpha(\rr)$, and binding energy $\epsilon_\mu(\rr)$. Here, $\Ecal^{(0)}=2\,{\rm eV}$ is the monolayer phosphorene band gap, extracted from \textit{ab initio} calculations \cite{tran2014layer}. All terms in the effective Hamiltonian \eqref{ExcHam} are given explicitly in the Supplementary Material. 

\begin{figure*}[t!]
    \centering
    \includegraphics{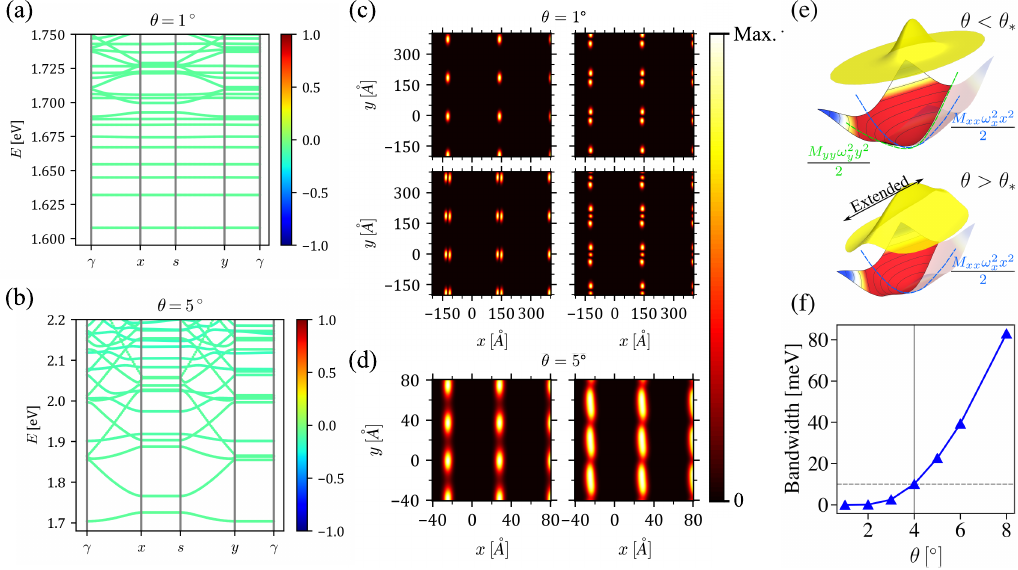}
    \caption{(a) Exciton miniband structure for twisted phosphorene bilayers encapsulated in hBN with $\theta=1^\circ$, and (b) $\theta=5^{\circ}$. The band colors indicate the intra- and interlayer exciton concent of each eigenstate, with blue (red) corresponding to a pure X (IX) state, and green to a maximally mixed hX state. (c) Average spatial distributions of the first four excitonic minibands for $\theta=1^{\circ}$, and (d) of the first two minibands for $\theta=5^{\circ}$. (e) Illustration of the moir\'e potential wells and their lowest energy states (offset for clarity), for $\theta < \theta_*$ and $\theta>\theta_*$. The approximating anisotropic harmonic oscillator potentials which confine the exciton are shown. (f) Lowest miniband width as a function of the twist angle. The $\theta_*=4^\circ$ threshold between the 0D and 1D CoM motion dimensionality regimes is indicated with a vertical line.}
    \label{fig:Fig3}
\end{figure*}

The moir\'e potential \eqref{ExcHam} can be diagonalized locally, treating the position $\rr$ as an adiabatic parameter. The spatial variation of the lowest energy level represents an effective potential landscape\cite{Ferreira_APL} for low-energy excitons propagating in the twisted phosphorene bilayer. Figure \ref{fig:Panel1}(c) shows that this potential landscape exhibits global minima with approximate $C_{2v}$ point symmetry at HH stacking regions, as well as saddle points at BA regions connecting neighboring potential wells along the $\hat{\mathbf{y}}$ direction. Below, we show that these potential wells are capable of fully localizing excitons for small twist angles ($\theta<4^\circ$), whereas at intermediate angles ($4^\circ < \theta \lesssim 10^\circ$) the excitons become delocalized exclusively along the $\hat{\mathbf{y}}$, or armchair direction.

\textit{Exciton minibands.}---The total effective Hamiltonian for excitons consists of the moir\'e potential \eqref{ExcHam}, plus the exciton kinetic energy
\begin{equation}
H_K(\QQ',\QQ)=\delta_{\QQ',\QQ} \mathbbm{1}_{4\times4}\frac{\hbar^2}{2}\QQ^T M_0^{-1} \QQ,
\end{equation}
with $\QQ$ the exciton CoM wave vector (treated here as a column vector), and $M_0^{-1}=\mathrm{diag}([m_x^c+m_x^v]^{-1},[m_y^c+m_y^v]^{-1})$ the anisotropic inverse exciton mass tensor, formed by the anisotropic conduction- and valence-band masses $m_x^c=1.12\,m_0$, $m_y^c=0.46\,m_0$, and $m_x^v=1.61\,m_0$, $m_y^v=0.23\,m_0$, respectively, with $m_0$ the free electron mass. We note that, for simplicity, we have neglected the effects of the relative layer rotation on the inverse mass tensors for intra- and interlayer excitons, thus introducing two sources of error into our calculations: Firstly, a total error below $3\%$ for both the X and IX CoM dispersions, and for the IX RM energies, at twist angles within the range of validity of our model. Secondly, the appearance of a perturbation that couples the IX CoM and RM degrees of freedom, much weaker than either the electron-hole interaction or the moir\'e potential, and which can thus be neglected as a first approximation. Further details can be found in the Supplementary Material.

We numerically diagonalized the total Hamiltonian using a zone-folding approach\cite{ruiz2019interlayer}: The moir\'e potential \eqref{ExcHam} mixes any X basis function \eqref{eq:Xbasis} at wave vector $\QQ$ with any IX basis function \eqref{eq:IXbasis} at wave vector $\QQ_{m,n} \equiv \QQ + \gG_{m,n}$. Since the mBZ is the Wigner-Seitz cell formed by the vectors $\gG_1$ and $\gG_2$, if we take $\QQ \in {\rm mBZ}$, all wave vectors $\QQ_{m,n}$ can be ``folded'' onto the mBZ, and relabeled as a state of superlattice wave vector $\QQ$ belonging to a so-called miniband $(m,n)$. In this scheme, the effective model becomes an independent eigenvalue problem for every $\QQ\in{\rm mBZ}$, which we solved numerically for a large but finite number of minibands, mutually coupled by the moir\'e potential \eqref{ExcHam}. Convergence to within a $1\,{\rm meV}$ tolerance was obtained for the lowest few energy eigenvalues for the range of indices $-12\leq m,n \leq 12$, corresponding to a total of 2500 basis states.

Figures \ref{fig:Fig3}(a) and \ref{fig:Fig3}(b) show the numerical moir\'e exciton miniband structures, computed for two representative twist angles: $\theta=1^\circ$ and $5^\circ$, respectively. For $\theta=1^\circ$, the lowest few minibands are completely flat, corresponding to Bloch states with vanishing group velocity. Intuition drawn from Fig.\ \ref{fig:Panel1}(c) tells us that these states are simply linear combinations of quantum-dot-like wave functions, strongly localized at HH stacking regions across the superlattice, with suppressed hopping between neighboring cells\cite{multifaceted}. This is verified in Fig.\ \ref{fig:Fig3}(c), which shows the mBZ-averaged exciton densities of the first four minibands of Fig.\ \ref{fig:Fig3}(a). In each case, the localization region of the states coincides with the minima of the potential landscape at HH stacking regions of the mSC [Fig.\ \ref{fig:Panel1}(c)]. The formation of multiple flat bands shows that, at small twist angles, such as $\theta=1^\circ$, the moir\'e potential wells are deep and wide enough to host several localized states, with spatial distributions reminiscent of the first few levels of a harmonic oscillator elongated in the $\hat{\mathbf{y}}$ direction.

All moir\'e exciton eigenstates obtained from our model are linear superpositions of X and IX states. In Figs.\ \ref{fig:Fig3}(a) and \ref{fig:Fig3}(b), we have color-coded the X and IX contents of each moir\'e exciton state, with blue (red) representing a pure X (IX) state, and green representing a maximally mixed state, known as a hybrid exciton (hX)\cite{alexeev2019resonantly,ruiz2019interlayer}. hXs are of wide interest for optoelectronics, as they combine the strong oscillator strength of Xs with the large electric dipole moment of IXs, making them simultaneously optically active and susceptible to out of plane electric fields. Our results of Fig.\ \ref{fig:Fig3}(a) indicate that all low-energy moir\'e excitons in a $\theta=1^\circ$ phosphorene bilayer are hXs, and thus both bright and tuneable.

\begin{figure}[t!]
    \centering
    \includegraphics[width=1.0\columnwidth]{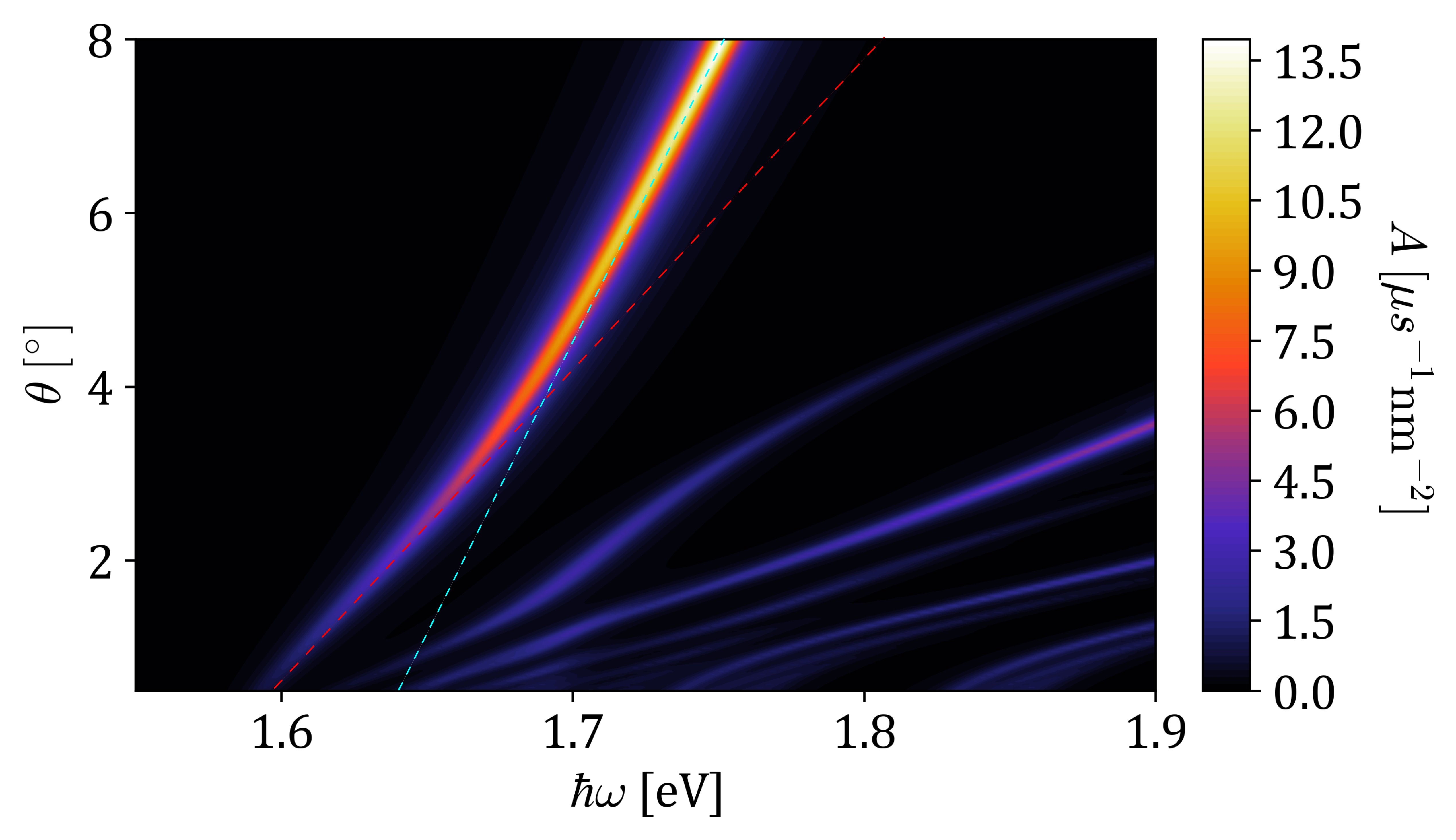}
    \caption{Absorption spectra for varying $\theta$ in the twisted phosphorene bilayer encapsulated in hBN. The linear fit for the first absorption line is shown for the $\theta<4^{\circ}$ regime (red) and for the $\theta>4^{\circ}$ regime (blue).}
    \label{fig:AbsSpec}
\end{figure}

\textit{One-dimensional moir\'e excitons.}---Figure \ref{fig:Fig3}(b) shows the computed miniband structure for a $\theta=5^{\circ}$ phosphorene bilayer, where the first few bands become dispersive in the $\hat{\mathbf{y}}$ (armchair) direction, while remaining flat along the $\hat{\mathbf{x}}$ (zig-zag) axis [see Fig.\ \ref{fig:ExcConfig}(a)]. This indicates the formation of moir\'e excitons delocalized in the former direction, but confined in the latter, representing a periodic array of quasi-1D states, reminiscent of quantum wires. To illustrate this, Fig.\ \ref{fig:Fig3}(d) shows the mBZ-averaged exciton densities of the first two minibands of Fig.\ \ref{fig:Fig3}(b). In addition, Fig.\ \ref{fig:Fig3}(b) also shows that all low energy moir\'e excitons remain maximally mixed hXs at intermediate twist angles.

Delocalization along the the armchair direction is mainly a consequence of the lighter exciton mass in the $\hat{\mathbf{y}}$ direction ($M_{xx}=2.73\,m_0$,\,$M_{yy}=0.69\,m_0$). The mSC shrinks as the twist angle increases, resulting in progressively narrower potential wells that eventually become unable to confine the moir\'e excitons. This occurs first along the $\hat{\mathbf{y}}$, or armchair axis, despite the larger width of the potential wells in that direction [Fig.\ \ref{fig:Panel1}(c)], due to the much lighter $M_{yy}$, as illustrated in Fig.\ \ref{fig:Fig3}(e). This is in stark contrast with the case of twisted transition-metal dichalcogenide bilayers, where, as the twist angle increases, moir\'e trapped excitons and carriers   become delocalized in the entire plane of the sample\cite{brem2020tunable,multifaceted}.

To quantitatively describe the crossover between the quantum-dot-like- (0D) and quantum-wire-like (1D) moir\'e exciton states, Fig.\ \ref{fig:Fig3}(f) shows the evolution of the lowest moir\'e exciton miniband width with the twist angle $\theta$. We propose a bandwidth of 10 meV as an empirical threshold, below (above) which a band can be considered flat (dispersive). The crossover angle $\theta_*$ between the 0D and 1D regimes can then be defined as the twist angle for which the lowest miniband width is 10 meV. This angle is found to be $\theta_{*}=4^{\circ}$, well within the range of validity of our model.

\textit{Moiré optical signatures.}---We have found direct optical signatures of the crossover between 0D- and 1D exciton states, experimentally accessible through absorption measurements. Details on our photo-absorption calculations can be found in the Supplementary Material. Figure \ref{fig:AbsSpec} shows the evolution of the absorption spectrum of $1s$ excitons in a twisted phosphorene bilayer, as a function of twist angle. The presence of moir\'e excitons can be inferred at first glance by the presence of multiple absorption lines\cite{alexeev2019resonantly,tran2019evidence,jin2019observation,seyler2019signatures} at energies close to that of the monolayer X state. These lines correspond to $\gamma$-point moir\'e excitons, and their oscillator strengths are dictated by the magnitude of the monolayer $\Gamma$-point exciton component in their wave functions. Focusing on the two leftmost absorption lines in Fig.\ \ref{fig:AbsSpec}, we see that both blueshift linearly with increasing twist angle, before they exhibit a sudden decrease in slope, treating $\theta$ as the abscissa. Figure \ref{fig:AbsSpec} shows that the twist angle at which the slope change occurs for the first absorption line coincides with our estimated dimensional crossover angle $\theta_*=4^\circ$. For the second line, this occurs for a twist angle slightly below $2^\circ$.

The absorption-line twist angle dependence can be understood in terms of the dimensional crossover of the moir\'e exciton states. In the 0D regime, the moir\'e exciton energies are well approximated by the zero-point energy of the confining potential wells, as illustrated on the top of Fig.\ \ref{fig:Fig3}(f). Since the confining potential has approximately rectangular ($C_{2v}$) symmetry, the zero-point energy is separable into two components,
\begin{equation}
\frac{\hbar\omega_x(\theta)}{2}+\frac{\hbar \omega_y(\theta)}{2} = \left( \frac{\hbar\omega_x^0}{2}+\frac{\hbar \omega_y^0}{2} \right) + (\sigma_x + \sigma_y)\theta,
\end{equation}
both of which increase linearly with slopes $\sigma_x,\sigma_y>0$ as the twist angle grows and the potential wells narrow. Passing the threshold angle $\theta_*$ into the 1D regime, the zero-point energy component $\hbar\omega_x(\theta)/2$ is replaced by the kinetic energy along the armchair direction, which vanishes for the lowest $\gamma$-point state. The energy of the optically active 1D moir\'e exciton then varies with $\theta$ as $\sigma_y\theta$, with a reduced slope $\sigma_y < \sigma_x+\sigma_y$, thus explaining the observed behavior.

\textit{Conclusions.}---We have predicted a dimensional crossover for moir\'e exciton states in twisted phosphorene bilayers, from quantum-dot-like (0D) to quantum-wire-like (1D) arrays, at an experimentally accessible twist angle of $\theta_*=4^\circ$. Our calculations show that the dimensionality regime can be identified experimentally by looking at the twist-angle dependence of the bilayer's optical absorption spectrum, which bears signatures of the dimensionality crossover. We have established that this crossover is driven by the large anisotropies of the carrier dispersions in the monolayer material, magnified by the moir\'e potential. As such, we expect analogous effects in other anisotropic 2D semiconductors, such as the group-IV monochalcogenides. Our results suggest that twisted bilayers of 2D semiconductors can realize versatile quantum many-body simulators, offering control over the system dimensionality. 

\text{}

\textit{Acknowledgments.} I.S.\ acknowledges financial support from CONACyT (M\'exico), through a \emph{Becas Nacionales} graduate scholarship, as well as from the University of Manchester's Dean's Doctoral Scholarship. D.A.R-T. acknowledges funding from PAPIIT-DGAPA-UNAM grant IA106523, and CONACyT (M\'exico) grants A1-S-14407 and 1564464. The authors would like to thank J.\ Guerrero-S\'anchez and F.\ Mireles for fruitful discussions at the beginning of this project.

\bibliography{references}

\begin{thebibliography}{40}%
\makeatletter
\providecommand \@ifxundefined [1]{%
 \@ifx{#1\undefined}
}%
\providecommand \@ifnum [1]{%
 \ifnum #1\expandafter \@firstoftwo
 \else \expandafter \@secondoftwo
 \fi
}%
\providecommand \@ifx [1]{%
 \ifx #1\expandafter \@firstoftwo
 \else \expandafter \@secondoftwo
 \fi
}%
\providecommand \natexlab [1]{#1}%
\providecommand \enquote  [1]{``#1''}%
\providecommand \bibnamefont  [1]{#1}%
\providecommand \bibfnamefont [1]{#1}%
\providecommand \citenamefont [1]{#1}%
\providecommand \href@noop [0]{\@secondoftwo}%
\providecommand \href [0]{\begingroup \@sanitize@url \@href}%
\providecommand \@href[1]{\@@startlink{#1}\@@href}%
\providecommand \@@href[1]{\endgroup#1\@@endlink}%
\providecommand \@sanitize@url [0]{\catcode `\\12\catcode `\$12\catcode
  `\&12\catcode `\#12\catcode `\^12\catcode `\_12\catcode `\%12\relax}%
\providecommand \@@startlink[1]{}%
\providecommand \@@endlink[0]{}%
\providecommand \url  [0]{\begingroup\@sanitize@url \@url }%
\providecommand \@url [1]{\endgroup\@href {#1}{\urlprefix }}%
\providecommand \urlprefix  [0]{URL }%
\providecommand \Eprint [0]{\href }%
\providecommand \doibase [0]{https://doi.org/}%
\providecommand \selectlanguage [0]{\@gobble}%
\providecommand \bibinfo  [0]{\@secondoftwo}%
\providecommand \bibfield  [0]{\@secondoftwo}%
\providecommand \translation [1]{[#1]}%
\providecommand \BibitemOpen [0]{}%
\providecommand \bibitemStop [0]{}%
\providecommand \bibitemNoStop [0]{.\EOS\space}%
\providecommand \EOS [0]{\spacefactor3000\relax}%
\providecommand \BibitemShut  [1]{\csname bibitem#1\endcsname}%
\let\auto@bib@innerbib\@empty
\bibitem [{\citenamefont {Soltero}\ \emph {et~al.}(2022)\citenamefont
  {Soltero}, \citenamefont {Guerrero-S\'anchez}, \citenamefont {Mireles},\ and\
  \citenamefont {Ruiz-Tijerina}}]{MoirePhos.PhysRevB.105.235421}%
  \BibitemOpen
  \bibfield  {author} {\bibinfo {author} {\bibfnamefont {I.}~\bibnamefont
  {Soltero}}, \bibinfo {author} {\bibfnamefont {J.}~\bibnamefont
  {Guerrero-S\'anchez}}, \bibinfo {author} {\bibfnamefont {F.}~\bibnamefont
  {Mireles}},\ and\ \bibinfo {author} {\bibfnamefont {D.~A.}\ \bibnamefont
  {Ruiz-Tijerina}},\ }\href {https://doi.org/10.1103/PhysRevB.105.235421}
  {\bibfield  {journal} {\bibinfo  {journal} {Physical Review B}\ }\textbf
  {\bibinfo {volume} {105}},\ \bibinfo {pages} {235421} (\bibinfo {year}
  {2022})}\BibitemShut {NoStop}%
\bibitem [{\citenamefont {Kennes}\ \emph {et~al.}(2021)\citenamefont {Kennes},
  \citenamefont {Claassen}, \citenamefont {Xian}, \citenamefont {Georges},
  \citenamefont {Millis}, \citenamefont {Hone}, \citenamefont {Dean},
  \citenamefont {Basov}, \citenamefont {Pasupathy},\ and\ \citenamefont
  {Rubio}}]{CMquantumsim2021}%
  \BibitemOpen
  \bibfield  {author} {\bibinfo {author} {\bibfnamefont {D.~M.}\ \bibnamefont
  {Kennes}}, \bibinfo {author} {\bibfnamefont {M.}~\bibnamefont {Claassen}},
  \bibinfo {author} {\bibfnamefont {L.}~\bibnamefont {Xian}}, \bibinfo {author}
  {\bibfnamefont {A.}~\bibnamefont {Georges}}, \bibinfo {author} {\bibfnamefont
  {A.~J.}\ \bibnamefont {Millis}}, \bibinfo {author} {\bibfnamefont
  {J.}~\bibnamefont {Hone}}, \bibinfo {author} {\bibfnamefont {C.~R.}\
  \bibnamefont {Dean}}, \bibinfo {author} {\bibfnamefont {D.}~\bibnamefont
  {Basov}}, \bibinfo {author} {\bibfnamefont {A.~N.}\ \bibnamefont
  {Pasupathy}},\ and\ \bibinfo {author} {\bibfnamefont {A.}~\bibnamefont
  {Rubio}},\ }\href@noop {} {\bibfield  {journal} {\bibinfo  {journal} {Nature
  Physics}\ }\textbf {\bibinfo {volume} {17}},\ \bibinfo {pages} {155}
  (\bibinfo {year} {2021})}\BibitemShut {NoStop}%
\bibitem [{\citenamefont {Wang}\ \emph {et~al.}(2020)\citenamefont {Wang},
  \citenamefont {Shih}, \citenamefont {Ghiotto}, \citenamefont {Xian},
  \citenamefont {Rhodes}, \citenamefont {Tan}, \citenamefont {Claassen},
  \citenamefont {Kennes}, \citenamefont {Bai}, \citenamefont {Kim} \emph
  {et~al.}}]{wang2020correlated}%
  \BibitemOpen
  \bibfield  {author} {\bibinfo {author} {\bibfnamefont {L.}~\bibnamefont
  {Wang}}, \bibinfo {author} {\bibfnamefont {E.-M.}\ \bibnamefont {Shih}},
  \bibinfo {author} {\bibfnamefont {A.}~\bibnamefont {Ghiotto}}, \bibinfo
  {author} {\bibfnamefont {L.}~\bibnamefont {Xian}}, \bibinfo {author}
  {\bibfnamefont {D.~A.}\ \bibnamefont {Rhodes}}, \bibinfo {author}
  {\bibfnamefont {C.}~\bibnamefont {Tan}}, \bibinfo {author} {\bibfnamefont
  {M.}~\bibnamefont {Claassen}}, \bibinfo {author} {\bibfnamefont {D.~M.}\
  \bibnamefont {Kennes}}, \bibinfo {author} {\bibfnamefont {Y.}~\bibnamefont
  {Bai}}, \bibinfo {author} {\bibfnamefont {B.}~\bibnamefont {Kim}}, \emph
  {et~al.},\ }\href@noop {} {\bibfield  {journal} {\bibinfo  {journal} {Nature
  Materials}\ }\textbf {\bibinfo {volume} {19}},\ \bibinfo {pages} {861}
  (\bibinfo {year} {2020})}\BibitemShut {NoStop}%
\bibitem [{\citenamefont {Tang}\ \emph {et~al.}(2020)\citenamefont {Tang},
  \citenamefont {Li}, \citenamefont {Li}, \citenamefont {Xu}, \citenamefont
  {Liu}, \citenamefont {Barmak}, \citenamefont {Watanabe}, \citenamefont
  {Taniguchi}, \citenamefont {MacDonald}, \citenamefont {Shan} \emph
  {et~al.}}]{tang2020simulation}%
  \BibitemOpen
  \bibfield  {author} {\bibinfo {author} {\bibfnamefont {Y.}~\bibnamefont
  {Tang}}, \bibinfo {author} {\bibfnamefont {L.}~\bibnamefont {Li}}, \bibinfo
  {author} {\bibfnamefont {T.}~\bibnamefont {Li}}, \bibinfo {author}
  {\bibfnamefont {Y.}~\bibnamefont {Xu}}, \bibinfo {author} {\bibfnamefont
  {S.}~\bibnamefont {Liu}}, \bibinfo {author} {\bibfnamefont {K.}~\bibnamefont
  {Barmak}}, \bibinfo {author} {\bibfnamefont {K.}~\bibnamefont {Watanabe}},
  \bibinfo {author} {\bibfnamefont {T.}~\bibnamefont {Taniguchi}}, \bibinfo
  {author} {\bibfnamefont {A.~H.}\ \bibnamefont {MacDonald}}, \bibinfo {author}
  {\bibfnamefont {J.}~\bibnamefont {Shan}}, \emph {et~al.},\ }\href@noop {}
  {\bibfield  {journal} {\bibinfo  {journal} {Nature}\ }\textbf {\bibinfo
  {volume} {579}},\ \bibinfo {pages} {353} (\bibinfo {year}
  {2020})}\BibitemShut {NoStop}%
\bibitem [{\citenamefont {Li}\ \emph {et~al.}(2021)\citenamefont {Li},
  \citenamefont {Li}, \citenamefont {Regan}, \citenamefont {Wang},
  \citenamefont {Zhao}, \citenamefont {Kahn}, \citenamefont {Yumigeta},
  \citenamefont {Blei}, \citenamefont {Taniguchi}, \citenamefont {Watanabe}
  \emph {et~al.}}]{li2021imaging}%
  \BibitemOpen
  \bibfield  {author} {\bibinfo {author} {\bibfnamefont {H.}~\bibnamefont
  {Li}}, \bibinfo {author} {\bibfnamefont {S.}~\bibnamefont {Li}}, \bibinfo
  {author} {\bibfnamefont {E.~C.}\ \bibnamefont {Regan}}, \bibinfo {author}
  {\bibfnamefont {D.}~\bibnamefont {Wang}}, \bibinfo {author} {\bibfnamefont
  {W.}~\bibnamefont {Zhao}}, \bibinfo {author} {\bibfnamefont {S.}~\bibnamefont
  {Kahn}}, \bibinfo {author} {\bibfnamefont {K.}~\bibnamefont {Yumigeta}},
  \bibinfo {author} {\bibfnamefont {M.}~\bibnamefont {Blei}}, \bibinfo {author}
  {\bibfnamefont {T.}~\bibnamefont {Taniguchi}}, \bibinfo {author}
  {\bibfnamefont {K.}~\bibnamefont {Watanabe}}, \emph {et~al.},\ }\href@noop {}
  {\bibfield  {journal} {\bibinfo  {journal} {Nature}\ }\textbf {\bibinfo
  {volume} {597}},\ \bibinfo {pages} {650} (\bibinfo {year}
  {2021})}\BibitemShut {NoStop}%
\bibitem [{\citenamefont {Wang}\ \emph {et~al.}(2022)\citenamefont {Wang},
  \citenamefont {Yu}, \citenamefont {Kwan}, \citenamefont {Jia}, \citenamefont
  {Lei}, \citenamefont {Klemenz}, \citenamefont {Cevallos}, \citenamefont
  {Singha}, \citenamefont {Devakul}, \citenamefont {Watanabe} \emph
  {et~al.}}]{wang2022one}%
  \BibitemOpen
  \bibfield  {author} {\bibinfo {author} {\bibfnamefont {P.}~\bibnamefont
  {Wang}}, \bibinfo {author} {\bibfnamefont {G.}~\bibnamefont {Yu}}, \bibinfo
  {author} {\bibfnamefont {Y.~H.}\ \bibnamefont {Kwan}}, \bibinfo {author}
  {\bibfnamefont {Y.}~\bibnamefont {Jia}}, \bibinfo {author} {\bibfnamefont
  {S.}~\bibnamefont {Lei}}, \bibinfo {author} {\bibfnamefont {S.}~\bibnamefont
  {Klemenz}}, \bibinfo {author} {\bibfnamefont {F.~A.}\ \bibnamefont
  {Cevallos}}, \bibinfo {author} {\bibfnamefont {R.}~\bibnamefont {Singha}},
  \bibinfo {author} {\bibfnamefont {T.}~\bibnamefont {Devakul}}, \bibinfo
  {author} {\bibfnamefont {K.}~\bibnamefont {Watanabe}}, \emph {et~al.},\
  }\href@noop {} {\bibfield  {journal} {\bibinfo  {journal} {Nature}\ }\textbf
  {\bibinfo {volume} {605}},\ \bibinfo {pages} {57} (\bibinfo {year}
  {2022})}\BibitemShut {NoStop}%
\bibitem [{\citenamefont {Ma}\ \emph {et~al.}(2021)\citenamefont {Ma},
  \citenamefont {Nguyen}, \citenamefont {Wang}, \citenamefont {Zeng},
  \citenamefont {Watanabe}, \citenamefont {Taniguchi}, \citenamefont
  {MacDonald}, \citenamefont {Mak},\ and\ \citenamefont
  {Shan}}]{ma2021strongly}%
  \BibitemOpen
  \bibfield  {author} {\bibinfo {author} {\bibfnamefont {L.}~\bibnamefont
  {Ma}}, \bibinfo {author} {\bibfnamefont {P.~X.}\ \bibnamefont {Nguyen}},
  \bibinfo {author} {\bibfnamefont {Z.}~\bibnamefont {Wang}}, \bibinfo {author}
  {\bibfnamefont {Y.}~\bibnamefont {Zeng}}, \bibinfo {author} {\bibfnamefont
  {K.}~\bibnamefont {Watanabe}}, \bibinfo {author} {\bibfnamefont
  {T.}~\bibnamefont {Taniguchi}}, \bibinfo {author} {\bibfnamefont {A.~H.}\
  \bibnamefont {MacDonald}}, \bibinfo {author} {\bibfnamefont {K.~F.}\
  \bibnamefont {Mak}},\ and\ \bibinfo {author} {\bibfnamefont {J.}~\bibnamefont
  {Shan}},\ }\href@noop {} {\bibfield  {journal} {\bibinfo  {journal} {Nature}\
  }\textbf {\bibinfo {volume} {598}},\ \bibinfo {pages} {585} (\bibinfo {year}
  {2021})}\BibitemShut {NoStop}%
\bibitem [{\citenamefont {Gu}\ \emph {et~al.}(2022)\citenamefont {Gu},
  \citenamefont {Ma}, \citenamefont {Liu}, \citenamefont {Watanabe},
  \citenamefont {Taniguchi}, \citenamefont {Hone}, \citenamefont {Shan},\ and\
  \citenamefont {Mak}}]{gu2022dipolar}%
  \BibitemOpen
  \bibfield  {author} {\bibinfo {author} {\bibfnamefont {J.}~\bibnamefont
  {Gu}}, \bibinfo {author} {\bibfnamefont {L.}~\bibnamefont {Ma}}, \bibinfo
  {author} {\bibfnamefont {S.}~\bibnamefont {Liu}}, \bibinfo {author}
  {\bibfnamefont {K.}~\bibnamefont {Watanabe}}, \bibinfo {author}
  {\bibfnamefont {T.}~\bibnamefont {Taniguchi}}, \bibinfo {author}
  {\bibfnamefont {J.~C.}\ \bibnamefont {Hone}}, \bibinfo {author}
  {\bibfnamefont {J.}~\bibnamefont {Shan}},\ and\ \bibinfo {author}
  {\bibfnamefont {K.~F.}\ \bibnamefont {Mak}},\ }\href@noop {} {\bibfield
  {journal} {\bibinfo  {journal} {Nature Physics}\ }\textbf {\bibinfo {volume}
  {18}},\ \bibinfo {pages} {395} (\bibinfo {year} {2022})}\BibitemShut
  {NoStop}%
\bibitem [{\citenamefont {Zeng}\ \emph {et~al.}(2023)\citenamefont {Zeng},
  \citenamefont {Xia}, \citenamefont {Dery}, \citenamefont {Watanabe},
  \citenamefont {Taniguchi}, \citenamefont {Shan},\ and\ \citenamefont
  {Mak}}]{zeng2023exciton}%
  \BibitemOpen
  \bibfield  {author} {\bibinfo {author} {\bibfnamefont {Y.}~\bibnamefont
  {Zeng}}, \bibinfo {author} {\bibfnamefont {Z.}~\bibnamefont {Xia}}, \bibinfo
  {author} {\bibfnamefont {R.}~\bibnamefont {Dery}}, \bibinfo {author}
  {\bibfnamefont {K.}~\bibnamefont {Watanabe}}, \bibinfo {author}
  {\bibfnamefont {T.}~\bibnamefont {Taniguchi}}, \bibinfo {author}
  {\bibfnamefont {J.}~\bibnamefont {Shan}},\ and\ \bibinfo {author}
  {\bibfnamefont {K.~F.}\ \bibnamefont {Mak}},\ }\href@noop {} {\bibfield
  {journal} {\bibinfo  {journal} {Nature Materials}\ ,\ \bibinfo {pages} {1}}
  (\bibinfo {year} {2023})}\BibitemShut {NoStop}%
\bibitem [{\citenamefont {Wu}\ \emph {et~al.}(2018)\citenamefont {Wu},
  \citenamefont {Lovorn}, \citenamefont {Tutuc},\ and\ \citenamefont
  {MacDonald}}]{wu2018hubbard}%
  \BibitemOpen
  \bibfield  {author} {\bibinfo {author} {\bibfnamefont {F.}~\bibnamefont
  {Wu}}, \bibinfo {author} {\bibfnamefont {T.}~\bibnamefont {Lovorn}}, \bibinfo
  {author} {\bibfnamefont {E.}~\bibnamefont {Tutuc}},\ and\ \bibinfo {author}
  {\bibfnamefont {A.~H.}\ \bibnamefont {MacDonald}},\ }\href@noop {} {\bibfield
   {journal} {\bibinfo  {journal} {Physical Review Letters}\ }\textbf {\bibinfo
  {volume} {121}},\ \bibinfo {pages} {026402} (\bibinfo {year}
  {2018})}\BibitemShut {NoStop}%
\bibitem [{\citenamefont {Angeli}\ and\ \citenamefont
  {MacDonald}(2021)}]{angeli2021gamma}%
  \BibitemOpen
  \bibfield  {author} {\bibinfo {author} {\bibfnamefont {M.}~\bibnamefont
  {Angeli}}\ and\ \bibinfo {author} {\bibfnamefont {A.~H.}\ \bibnamefont
  {MacDonald}},\ }\href@noop {} {\bibfield  {journal} {\bibinfo  {journal}
  {Proceedings of the National Academy of Sciences}\ }\textbf {\bibinfo
  {volume} {118}},\ \bibinfo {pages} {e2021826118} (\bibinfo {year}
  {2021})}\BibitemShut {NoStop}%
\bibitem [{\citenamefont {Magorrian}\ \emph {et~al.}(2021)\citenamefont
  {Magorrian}, \citenamefont {Enaldiev}, \citenamefont {Z\'olyomi},
  \citenamefont {Ferreira}, \citenamefont {Fal'ko},\ and\ \citenamefont
  {Ruiz-Tijerina}}]{multifaceted}%
  \BibitemOpen
  \bibfield  {author} {\bibinfo {author} {\bibfnamefont {S.~J.}\ \bibnamefont
  {Magorrian}}, \bibinfo {author} {\bibfnamefont {V.~V.}\ \bibnamefont
  {Enaldiev}}, \bibinfo {author} {\bibfnamefont {V.}~\bibnamefont {Z\'olyomi}},
  \bibinfo {author} {\bibfnamefont {F.}~\bibnamefont {Ferreira}}, \bibinfo
  {author} {\bibfnamefont {V.~I.}\ \bibnamefont {Fal'ko}},\ and\ \bibinfo
  {author} {\bibfnamefont {D.~A.}\ \bibnamefont {Ruiz-Tijerina}},\ }\href
  {https://doi.org/10.1103/PhysRevB.104.125440} {\bibfield  {journal} {\bibinfo
   {journal} {Phys. Rev. B}\ }\textbf {\bibinfo {volume} {104}},\ \bibinfo
  {pages} {125440} (\bibinfo {year} {2021})}\BibitemShut {NoStop}%
\bibitem [{\citenamefont {Kennes}\ \emph {et~al.}(2020)\citenamefont {Kennes},
  \citenamefont {Xian}, \citenamefont {Claassen},\ and\ \citenamefont
  {Rubio}}]{RubioGeSe}%
  \BibitemOpen
  \bibfield  {author} {\bibinfo {author} {\bibfnamefont {D.~M.}\ \bibnamefont
  {Kennes}}, \bibinfo {author} {\bibfnamefont {L.}~\bibnamefont {Xian}},
  \bibinfo {author} {\bibfnamefont {M.}~\bibnamefont {Claassen}},\ and\
  \bibinfo {author} {\bibfnamefont {A.}~\bibnamefont {Rubio}},\ }\href@noop {}
  {\bibfield  {journal} {\bibinfo  {journal} {Nature communications}\ }\textbf
  {\bibinfo {volume} {11}},\ \bibinfo {pages} {1124} (\bibinfo {year}
  {2020})}\BibitemShut {NoStop}%
\bibitem [{\citenamefont {Fujimoto}\ and\ \citenamefont
  {Kariyado}(2021)}]{KariyadoGeSe}%
  \BibitemOpen
  \bibfield  {author} {\bibinfo {author} {\bibfnamefont {M.}~\bibnamefont
  {Fujimoto}}\ and\ \bibinfo {author} {\bibfnamefont {T.}~\bibnamefont
  {Kariyado}},\ }\href {https://doi.org/10.1103/PhysRevB.104.125427} {\bibfield
   {journal} {\bibinfo  {journal} {Phys. Rev. B}\ }\textbf {\bibinfo {volume}
  {104}},\ \bibinfo {pages} {125427} (\bibinfo {year} {2021})}\BibitemShut
  {NoStop}%
\bibitem [{\citenamefont {Giamarchi}(2003)}]{giamarchi2003quantum}%
  \BibitemOpen
  \bibfield  {author} {\bibinfo {author} {\bibfnamefont {T.}~\bibnamefont
  {Giamarchi}},\ }\href@noop {} {\emph {\bibinfo {title} {Quantum physics in
  one dimension}}},\ Vol.\ \bibinfo {volume} {121}\ (\bibinfo  {publisher}
  {Clarendon press},\ \bibinfo {year} {2003})\BibitemShut {NoStop}%
\bibitem [{\citenamefont {Guo}\ \emph {et~al.}(2023)\citenamefont {Guo},
  \citenamefont {Zhang},\ and\ \citenamefont {Lu}}]{guo2023pseudo}%
  \BibitemOpen
  \bibfield  {author} {\bibinfo {author} {\bibfnamefont {H.}~\bibnamefont
  {Guo}}, \bibinfo {author} {\bibfnamefont {X.}~\bibnamefont {Zhang}},\ and\
  \bibinfo {author} {\bibfnamefont {G.}~\bibnamefont {Lu}},\ }\href@noop {}
  {\bibfield  {journal} {\bibinfo  {journal} {Science Advances}\ }\textbf
  {\bibinfo {volume} {9}},\ \bibinfo {pages} {eadi5404} (\bibinfo {year}
  {2023})}\BibitemShut {NoStop}%
\bibitem [{\citenamefont {Bai}\ \emph {et~al.}(2020)\citenamefont {Bai},
  \citenamefont {Zhou}, \citenamefont {Wang}, \citenamefont {Wu}, \citenamefont
  {McGilly}, \citenamefont {Halbertal}, \citenamefont {Lo}, \citenamefont
  {Liu}, \citenamefont {Ardelean}, \citenamefont {Rivera} \emph
  {et~al.}}]{bai2020excitons}%
  \BibitemOpen
  \bibfield  {author} {\bibinfo {author} {\bibfnamefont {Y.}~\bibnamefont
  {Bai}}, \bibinfo {author} {\bibfnamefont {L.}~\bibnamefont {Zhou}}, \bibinfo
  {author} {\bibfnamefont {J.}~\bibnamefont {Wang}}, \bibinfo {author}
  {\bibfnamefont {W.}~\bibnamefont {Wu}}, \bibinfo {author} {\bibfnamefont
  {L.~J.}\ \bibnamefont {McGilly}}, \bibinfo {author} {\bibfnamefont
  {D.}~\bibnamefont {Halbertal}}, \bibinfo {author} {\bibfnamefont {C.~F.~B.}\
  \bibnamefont {Lo}}, \bibinfo {author} {\bibfnamefont {F.}~\bibnamefont
  {Liu}}, \bibinfo {author} {\bibfnamefont {J.}~\bibnamefont {Ardelean}},
  \bibinfo {author} {\bibfnamefont {P.}~\bibnamefont {Rivera}}, \emph
  {et~al.},\ }\href@noop {} {\bibfield  {journal} {\bibinfo  {journal} {Nature
  Materials}\ }\textbf {\bibinfo {volume} {19}},\ \bibinfo {pages} {1068}
  (\bibinfo {year} {2020})}\BibitemShut {NoStop}%
\bibitem [{\citenamefont {Yang}\ \emph {et~al.}(2017)\citenamefont {Yang},
  \citenamefont {Chen}, \citenamefont {Zheng}, \citenamefont {Sun},
  \citenamefont {Dai}, \citenamefont {Guan}, \citenamefont {Yuan},\ and\
  \citenamefont {Pan}}]{coldatoms1}%
  \BibitemOpen
  \bibfield  {author} {\bibinfo {author} {\bibfnamefont {B.}~\bibnamefont
  {Yang}}, \bibinfo {author} {\bibfnamefont {Y.-Y.}\ \bibnamefont {Chen}},
  \bibinfo {author} {\bibfnamefont {Y.-G.}\ \bibnamefont {Zheng}}, \bibinfo
  {author} {\bibfnamefont {H.}~\bibnamefont {Sun}}, \bibinfo {author}
  {\bibfnamefont {H.-N.}\ \bibnamefont {Dai}}, \bibinfo {author} {\bibfnamefont
  {X.-W.}\ \bibnamefont {Guan}}, \bibinfo {author} {\bibfnamefont {Z.-S.}\
  \bibnamefont {Yuan}},\ and\ \bibinfo {author} {\bibfnamefont {J.-W.}\
  \bibnamefont {Pan}},\ }\href {https://doi.org/10.1103/PhysRevLett.119.165701}
  {\bibfield  {journal} {\bibinfo  {journal} {Phys. Rev. Lett.}\ }\textbf
  {\bibinfo {volume} {119}},\ \bibinfo {pages} {165701} (\bibinfo {year}
  {2017})}\BibitemShut {NoStop}%
\bibitem [{\citenamefont {Lake}\ \emph {et~al.}(2021)\citenamefont {Lake},
  \citenamefont {Senthil},\ and\ \citenamefont
  {Vishwanath}}]{BoseLuttingerVishwanath}%
  \BibitemOpen
  \bibfield  {author} {\bibinfo {author} {\bibfnamefont {E.}~\bibnamefont
  {Lake}}, \bibinfo {author} {\bibfnamefont {T.}~\bibnamefont {Senthil}},\ and\
  \bibinfo {author} {\bibfnamefont {A.}~\bibnamefont {Vishwanath}},\ }\href
  {https://doi.org/10.1103/PhysRevB.104.014517} {\bibfield  {journal} {\bibinfo
   {journal} {Phys. Rev. B}\ }\textbf {\bibinfo {volume} {104}},\ \bibinfo
  {pages} {014517} (\bibinfo {year} {2021})}\BibitemShut {NoStop}%
\bibitem [{\citenamefont {Yao}\ \emph {et~al.}(2023)\citenamefont {Yao},
  \citenamefont {Pizzino},\ and\ \citenamefont {Giamarchi}}]{giamarchi1d2d}%
  \BibitemOpen
  \bibfield  {author} {\bibinfo {author} {\bibfnamefont {H.}~\bibnamefont
  {Yao}}, \bibinfo {author} {\bibfnamefont {L.}~\bibnamefont {Pizzino}},\ and\
  \bibinfo {author} {\bibfnamefont {T.}~\bibnamefont {Giamarchi}},\ }\bibfield
  {journal} {\bibinfo  {journal} {SciPost Phys.}\ }\href
  {https://doi.org/10.21468/SciPostPhys.15.2.050}
  {10.21468/SciPostPhys.15.2.050} (\bibinfo {year} {2023})\BibitemShut
  {NoStop}%
\bibitem [{\citenamefont {Geick}\ \emph {et~al.}(1966)\citenamefont {Geick},
  \citenamefont {Perry},\ and\ \citenamefont {Rupprecht}}]{geick1966normal}%
  \BibitemOpen
  \bibfield  {author} {\bibinfo {author} {\bibfnamefont {R.}~\bibnamefont
  {Geick}}, \bibinfo {author} {\bibfnamefont {C.}~\bibnamefont {Perry}},\ and\
  \bibinfo {author} {\bibfnamefont {G.}~\bibnamefont {Rupprecht}},\ }\href@noop
  {} {\bibfield  {journal} {\bibinfo  {journal} {Physical Review}\ }\textbf
  {\bibinfo {volume} {146}},\ \bibinfo {pages} {543} (\bibinfo {year}
  {1966})}\BibitemShut {NoStop}%
\bibitem [{\citenamefont {Laturia}\ \emph {et~al.}(2018)\citenamefont
  {Laturia}, \citenamefont {Van~de Put},\ and\ \citenamefont
  {Vandenberghe}}]{laturia2018dielectric}%
  \BibitemOpen
  \bibfield  {author} {\bibinfo {author} {\bibfnamefont {A.}~\bibnamefont
  {Laturia}}, \bibinfo {author} {\bibfnamefont {M.~L.}\ \bibnamefont {Van~de
  Put}},\ and\ \bibinfo {author} {\bibfnamefont {W.~G.}\ \bibnamefont
  {Vandenberghe}},\ }\href@noop {} {\bibfield  {journal} {\bibinfo  {journal}
  {npj 2D Materials and Applications}\ }\textbf {\bibinfo {volume} {2}},\
  \bibinfo {pages} {1} (\bibinfo {year} {2018})}\BibitemShut {NoStop}%
\bibitem [{\citenamefont {Danovich}\ \emph {et~al.}(2018)\citenamefont
  {Danovich}, \citenamefont {Ruiz-Tijerina}, \citenamefont {Hunt},
  \citenamefont {Szyniszewski}, \citenamefont {Drummond},\ and\ \citenamefont
  {Fal'ko}}]{danovich2018localized}%
  \BibitemOpen
  \bibfield  {author} {\bibinfo {author} {\bibfnamefont {M.}~\bibnamefont
  {Danovich}}, \bibinfo {author} {\bibfnamefont {D.~A.}\ \bibnamefont
  {Ruiz-Tijerina}}, \bibinfo {author} {\bibfnamefont {R.~J.}\ \bibnamefont
  {Hunt}}, \bibinfo {author} {\bibfnamefont {M.}~\bibnamefont {Szyniszewski}},
  \bibinfo {author} {\bibfnamefont {N.~D.}\ \bibnamefont {Drummond}},\ and\
  \bibinfo {author} {\bibfnamefont {V.~I.}\ \bibnamefont {Fal'ko}},\
  }\href@noop {} {\bibfield  {journal} {\bibinfo  {journal} {Physical Review
  B}\ }\textbf {\bibinfo {volume} {97}},\ \bibinfo {pages} {195452} (\bibinfo
  {year} {2018})}\BibitemShut {NoStop}%
\bibitem [{\citenamefont {Viner}\ \emph {et~al.}(2021)\citenamefont {Viner},
  \citenamefont {McDonnell}, \citenamefont {Ruiz-Tijerina}, \citenamefont
  {Rivera}, \citenamefont {Xu}, \citenamefont {Fal’ko},\ and\ \citenamefont
  {Smith}}]{viner2021excited}%
  \BibitemOpen
  \bibfield  {author} {\bibinfo {author} {\bibfnamefont {J.~J.}\ \bibnamefont
  {Viner}}, \bibinfo {author} {\bibfnamefont {L.~P.}\ \bibnamefont
  {McDonnell}}, \bibinfo {author} {\bibfnamefont {D.~A.}\ \bibnamefont
  {Ruiz-Tijerina}}, \bibinfo {author} {\bibfnamefont {P.}~\bibnamefont
  {Rivera}}, \bibinfo {author} {\bibfnamefont {X.}~\bibnamefont {Xu}}, \bibinfo
  {author} {\bibfnamefont {V.~I.}\ \bibnamefont {Fal’ko}},\ and\ \bibinfo
  {author} {\bibfnamefont {D.~C.}\ \bibnamefont {Smith}},\ }\href@noop {}
  {\bibfield  {journal} {\bibinfo  {journal} {2D Materials}\ }\textbf {\bibinfo
  {volume} {8}},\ \bibinfo {pages} {035047} (\bibinfo {year}
  {2021})}\BibitemShut {NoStop}%
\bibitem [{\citenamefont {Henriques}\ and\ \citenamefont
  {Peres}(2020)}]{henriques2020excitons}%
  \BibitemOpen
  \bibfield  {author} {\bibinfo {author} {\bibfnamefont {J.}~\bibnamefont
  {Henriques}}\ and\ \bibinfo {author} {\bibfnamefont {N.}~\bibnamefont
  {Peres}},\ }\href@noop {} {\bibfield  {journal} {\bibinfo  {journal}
  {Physical Review B}\ }\textbf {\bibinfo {volume} {101}},\ \bibinfo {pages}
  {035406} (\bibinfo {year} {2020})}\BibitemShut {NoStop}%
\bibitem [{Note1()}]{Note1}%
  \BibitemOpen
  \bibinfo {note} {We have also considered the RM wave functions as
  adiabatically depending on the local stacking configuration: $X[\protect \bm
  {\rho },\protect \mathbf {r}_0(\protect \mathbf {r})]$ and $Y[\protect \bm
  {\rho },\protect \mathbf {r}_0(\protect \mathbf {r})]$. The $\protect \mathbf
  {r}_0$-dependence of these functions introduces an additional spatial
  dependence to the matrix elements $ \leavevmode@ifvmode {\setbox \z@ \hbox
  {\mathsurround \z@ $\nulldelimiterspace \z@ \left <\vcenter to\@ne \big@size
  {}\right .$}\box \z@ } {\protect \rm X}_{\lambda ',n'}(\protect \mathbf {Q}')
  \leavevmode@ifvmode {\setbox \z@ \hbox {\mathsurround \z@
  $\nulldelimiterspace \z@ \left |\vcenter to\@ne \big@size {}\right .$}\box
  \z@ } H_{\protect \rm exc} \leavevmode@ifvmode {\setbox \z@ \hbox
  {\mathsurround \z@ $\nulldelimiterspace \z@ \left |\vcenter to\@ne \big@size
  {}\right .$}\box \z@ } {\protect \rm IX}_{\lambda ,n}^{\bar {\lambda
  }}(\protect \mathbf {Q}) \leavevmode@ifvmode {\setbox \z@ \hbox
  {\mathsurround \z@ $\nulldelimiterspace \z@ \left >\vcenter to\@ne \big@size
  {}\right .$}\box \z@ } $. We have numerically determined that this variation
  is $<1\%$, and thus negligible.}\BibitemShut {Stop}%
\bibitem [{\citenamefont {Griffin}\ and\ \citenamefont
  {Wheeler}(1957)}]{griffin1957collective}%
  \BibitemOpen
  \bibfield  {author} {\bibinfo {author} {\bibfnamefont {J.~J.}\ \bibnamefont
  {Griffin}}\ and\ \bibinfo {author} {\bibfnamefont {J.~A.}\ \bibnamefont
  {Wheeler}},\ }\href@noop {} {\bibfield  {journal} {\bibinfo  {journal}
  {Physical Review}\ }\textbf {\bibinfo {volume} {108}},\ \bibinfo {pages}
  {311} (\bibinfo {year} {1957})}\BibitemShut {NoStop}%
\bibitem [{\citenamefont {Henriques}\ \emph {et~al.}(2019)\citenamefont
  {Henriques}, \citenamefont {Ventura}, \citenamefont {Fernandes},\ and\
  \citenamefont {Peres}}]{henriques2019optical}%
  \BibitemOpen
  \bibfield  {author} {\bibinfo {author} {\bibfnamefont {J.}~\bibnamefont
  {Henriques}}, \bibinfo {author} {\bibfnamefont {G.}~\bibnamefont {Ventura}},
  \bibinfo {author} {\bibfnamefont {C.}~\bibnamefont {Fernandes}},\ and\
  \bibinfo {author} {\bibfnamefont {N.}~\bibnamefont {Peres}},\ }\href@noop {}
  {\bibfield  {journal} {\bibinfo  {journal} {Journal of Physics: Condensed
  Matter}\ }\textbf {\bibinfo {volume} {32}},\ \bibinfo {pages} {025304}
  (\bibinfo {year} {2019})}\BibitemShut {NoStop}%
\bibitem [{\citenamefont {Ruiz-Tijerina}\ \emph {et~al.}(2020)\citenamefont
  {Ruiz-Tijerina}, \citenamefont {Soltero},\ and\ \citenamefont
  {Mireles}}]{ruiz2020theory}%
  \BibitemOpen
  \bibfield  {author} {\bibinfo {author} {\bibfnamefont {D.~A.}\ \bibnamefont
  {Ruiz-Tijerina}}, \bibinfo {author} {\bibfnamefont {I.}~\bibnamefont
  {Soltero}},\ and\ \bibinfo {author} {\bibfnamefont {F.}~\bibnamefont
  {Mireles}},\ }\href@noop {} {\bibfield  {journal} {\bibinfo  {journal}
  {Physical Review B}\ }\textbf {\bibinfo {volume} {102}},\ \bibinfo {pages}
  {195403} (\bibinfo {year} {2020})}\BibitemShut {NoStop}%
\bibitem [{Note2()}]{Note2}%
  \BibitemOpen
  \bibinfo {note} {The lowest X and IX states transform as the $A_1$
  irreducible representation of the symmetry group $C_{2v}$ of the RM
  Hamiltonian, and look like $1s$ hydrogenic states elongated in the $y$
  direction (see Supplementary Material), justifying the label $1s$. The
  eventual importance of, \protect \emph {e.g.}, $2p$ or $2s$ excited states
  for the $1s$ moir\'e exciton band structures is determined by the $2p-2s$ and
  $2s-1s$ wave function overlaps, which we estimate to be at least one order of
  magnitude smaller than any $1s-1s$ overlap. Moreover, the oscillator strength
  of the $2s$ intralayer exciton is also estimated to be much weaker than that
  of its $1s$ counterpart, such that it can be neglected in the optical
  spectrum.}\BibitemShut {Stop}%
\bibitem [{\citenamefont {Tran}\ \emph {et~al.}(2014)\citenamefont {Tran},
  \citenamefont {Soklaski}, \citenamefont {Liang},\ and\ \citenamefont
  {Yang}}]{tran2014layer}%
  \BibitemOpen
  \bibfield  {author} {\bibinfo {author} {\bibfnamefont {V.}~\bibnamefont
  {Tran}}, \bibinfo {author} {\bibfnamefont {R.}~\bibnamefont {Soklaski}},
  \bibinfo {author} {\bibfnamefont {Y.}~\bibnamefont {Liang}},\ and\ \bibinfo
  {author} {\bibfnamefont {L.}~\bibnamefont {Yang}},\ }\href@noop {} {\bibfield
   {journal} {\bibinfo  {journal} {Physical Review B}\ }\textbf {\bibinfo
  {volume} {89}},\ \bibinfo {pages} {235319} (\bibinfo {year}
  {2014})}\BibitemShut {NoStop}%
\bibitem [{\citenamefont {Ferreira}\ \emph {et~al.}(2021)\citenamefont
  {Ferreira}, \citenamefont {Magorrian}, \citenamefont {Enaldiev},
  \citenamefont {Ruiz-Tijerina},\ and\ \citenamefont {Fal'ko}}]{Ferreira_APL}%
  \BibitemOpen
  \bibfield  {author} {\bibinfo {author} {\bibfnamefont {F.}~\bibnamefont
  {Ferreira}}, \bibinfo {author} {\bibfnamefont {S.~J.}\ \bibnamefont
  {Magorrian}}, \bibinfo {author} {\bibfnamefont {V.~V.}\ \bibnamefont
  {Enaldiev}}, \bibinfo {author} {\bibfnamefont {D.~A.}\ \bibnamefont
  {Ruiz-Tijerina}},\ and\ \bibinfo {author} {\bibfnamefont {V.~I.}\
  \bibnamefont {Fal'ko}},\ }\href {https://doi.org/10.1063/5.0048884}
  {\bibfield  {journal} {\bibinfo  {journal} {Appl. Phys. Lett.}\ }\textbf
  {\bibinfo {volume} {118}},\ \bibinfo {pages} {241602} (\bibinfo {year}
  {2021})},\ \Eprint {https://arxiv.org/abs/https://doi.org/10.1063/5.0048884}
  {https://doi.org/10.1063/5.0048884} \BibitemShut {NoStop}%
\bibitem [{\citenamefont {Ruiz-Tijerina}\ and\ \citenamefont
  {Fal'ko}(2019)}]{ruiz2019interlayer}%
  \BibitemOpen
  \bibfield  {author} {\bibinfo {author} {\bibfnamefont {D.~A.}\ \bibnamefont
  {Ruiz-Tijerina}}\ and\ \bibinfo {author} {\bibfnamefont {V.~I.}\ \bibnamefont
  {Fal'ko}},\ }\href@noop {} {\bibfield  {journal} {\bibinfo  {journal}
  {Physical Review B}\ }\textbf {\bibinfo {volume} {99}},\ \bibinfo {pages}
  {125424} (\bibinfo {year} {2019})}\BibitemShut {NoStop}%
\bibitem [{\citenamefont {Alexeev}\ \emph {et~al.}(2019)\citenamefont
  {Alexeev}, \citenamefont {Ruiz-Tijerina}, \citenamefont {Danovich},
  \citenamefont {Hamer}, \citenamefont {Terry}, \citenamefont {Nayak},
  \citenamefont {Ahn}, \citenamefont {Pak}, \citenamefont {Lee}, \citenamefont
  {Sohn} \emph {et~al.}}]{alexeev2019resonantly}%
  \BibitemOpen
  \bibfield  {author} {\bibinfo {author} {\bibfnamefont {E.~M.}\ \bibnamefont
  {Alexeev}}, \bibinfo {author} {\bibfnamefont {D.~A.}\ \bibnamefont
  {Ruiz-Tijerina}}, \bibinfo {author} {\bibfnamefont {M.}~\bibnamefont
  {Danovich}}, \bibinfo {author} {\bibfnamefont {M.~J.}\ \bibnamefont {Hamer}},
  \bibinfo {author} {\bibfnamefont {D.~J.}\ \bibnamefont {Terry}}, \bibinfo
  {author} {\bibfnamefont {P.~K.}\ \bibnamefont {Nayak}}, \bibinfo {author}
  {\bibfnamefont {S.}~\bibnamefont {Ahn}}, \bibinfo {author} {\bibfnamefont
  {S.}~\bibnamefont {Pak}}, \bibinfo {author} {\bibfnamefont {J.}~\bibnamefont
  {Lee}}, \bibinfo {author} {\bibfnamefont {J.~I.}\ \bibnamefont {Sohn}}, \emph
  {et~al.},\ }\href@noop {} {\bibfield  {journal} {\bibinfo  {journal}
  {Nature}\ }\textbf {\bibinfo {volume} {567}},\ \bibinfo {pages} {81}
  (\bibinfo {year} {2019})}\BibitemShut {NoStop}%
\bibitem [{\citenamefont {Brem}\ \emph {et~al.}(2020)\citenamefont {Brem},
  \citenamefont {Linder\"alv}, \citenamefont {Erhart},\ and\ \citenamefont
  {Malic}}]{brem2020tunable}%
  \BibitemOpen
  \bibfield  {author} {\bibinfo {author} {\bibfnamefont {S.}~\bibnamefont
  {Brem}}, \bibinfo {author} {\bibfnamefont {C.}~\bibnamefont {Linder\"alv}},
  \bibinfo {author} {\bibfnamefont {P.}~\bibnamefont {Erhart}},\ and\ \bibinfo
  {author} {\bibfnamefont {E.}~\bibnamefont {Malic}},\ }\href@noop {}
  {\bibfield  {journal} {\bibinfo  {journal} {Nano letters}\ }\textbf {\bibinfo
  {volume} {20}},\ \bibinfo {pages} {8534} (\bibinfo {year}
  {2020})}\BibitemShut {NoStop}%
\bibitem [{\citenamefont {Tran}\ \emph {et~al.}(2019)\citenamefont {Tran},
  \citenamefont {Moody}, \citenamefont {Wu}, \citenamefont {Lu}, \citenamefont
  {Choi}, \citenamefont {Kim}, \citenamefont {Rai}, \citenamefont {Sanchez},
  \citenamefont {Quan}, \citenamefont {Singh} \emph
  {et~al.}}]{tran2019evidence}%
  \BibitemOpen
  \bibfield  {author} {\bibinfo {author} {\bibfnamefont {K.}~\bibnamefont
  {Tran}}, \bibinfo {author} {\bibfnamefont {G.}~\bibnamefont {Moody}},
  \bibinfo {author} {\bibfnamefont {F.}~\bibnamefont {Wu}}, \bibinfo {author}
  {\bibfnamefont {X.}~\bibnamefont {Lu}}, \bibinfo {author} {\bibfnamefont
  {J.}~\bibnamefont {Choi}}, \bibinfo {author} {\bibfnamefont {K.}~\bibnamefont
  {Kim}}, \bibinfo {author} {\bibfnamefont {A.}~\bibnamefont {Rai}}, \bibinfo
  {author} {\bibfnamefont {D.~A.}\ \bibnamefont {Sanchez}}, \bibinfo {author}
  {\bibfnamefont {J.}~\bibnamefont {Quan}}, \bibinfo {author} {\bibfnamefont
  {A.}~\bibnamefont {Singh}}, \emph {et~al.},\ }\href@noop {} {\bibfield
  {journal} {\bibinfo  {journal} {Nature}\ }\textbf {\bibinfo {volume} {567}},\
  \bibinfo {pages} {71} (\bibinfo {year} {2019})}\BibitemShut {NoStop}%
\bibitem [{\citenamefont {Jin}\ \emph {et~al.}(2019)\citenamefont {Jin},
  \citenamefont {Regan}, \citenamefont {Yan}, \citenamefont {Iqbal
  Bakti~Utama}, \citenamefont {Wang}, \citenamefont {Zhao}, \citenamefont
  {Qin}, \citenamefont {Yang}, \citenamefont {Zheng}, \citenamefont {Shi} \emph
  {et~al.}}]{jin2019observation}%
  \BibitemOpen
  \bibfield  {author} {\bibinfo {author} {\bibfnamefont {C.}~\bibnamefont
  {Jin}}, \bibinfo {author} {\bibfnamefont {E.~C.}\ \bibnamefont {Regan}},
  \bibinfo {author} {\bibfnamefont {A.}~\bibnamefont {Yan}}, \bibinfo {author}
  {\bibfnamefont {M.}~\bibnamefont {Iqbal Bakti~Utama}}, \bibinfo {author}
  {\bibfnamefont {D.}~\bibnamefont {Wang}}, \bibinfo {author} {\bibfnamefont
  {S.}~\bibnamefont {Zhao}}, \bibinfo {author} {\bibfnamefont {Y.}~\bibnamefont
  {Qin}}, \bibinfo {author} {\bibfnamefont {S.}~\bibnamefont {Yang}}, \bibinfo
  {author} {\bibfnamefont {Z.}~\bibnamefont {Zheng}}, \bibinfo {author}
  {\bibfnamefont {S.}~\bibnamefont {Shi}}, \emph {et~al.},\ }\href@noop {}
  {\bibfield  {journal} {\bibinfo  {journal} {Nature}\ }\textbf {\bibinfo
  {volume} {567}},\ \bibinfo {pages} {76} (\bibinfo {year} {2019})}\BibitemShut
  {NoStop}%
\bibitem [{\citenamefont {Seyler}\ \emph {et~al.}(2019)\citenamefont {Seyler},
  \citenamefont {Rivera}, \citenamefont {Yu}, \citenamefont {Wilson},
  \citenamefont {Ray}, \citenamefont {Mandrus}, \citenamefont {Yan},
  \citenamefont {Yao},\ and\ \citenamefont {Xu}}]{seyler2019signatures}%
  \BibitemOpen
  \bibfield  {author} {\bibinfo {author} {\bibfnamefont {K.~L.}\ \bibnamefont
  {Seyler}}, \bibinfo {author} {\bibfnamefont {P.}~\bibnamefont {Rivera}},
  \bibinfo {author} {\bibfnamefont {H.}~\bibnamefont {Yu}}, \bibinfo {author}
  {\bibfnamefont {N.~P.}\ \bibnamefont {Wilson}}, \bibinfo {author}
  {\bibfnamefont {E.~L.}\ \bibnamefont {Ray}}, \bibinfo {author} {\bibfnamefont
  {D.~G.}\ \bibnamefont {Mandrus}}, \bibinfo {author} {\bibfnamefont
  {J.}~\bibnamefont {Yan}}, \bibinfo {author} {\bibfnamefont {W.}~\bibnamefont
  {Yao}},\ and\ \bibinfo {author} {\bibfnamefont {X.}~\bibnamefont {Xu}},\
  }\href@noop {} {\bibfield  {journal} {\bibinfo  {journal} {Nature}\ }\textbf
  {\bibinfo {volume} {567}},\ \bibinfo {pages} {66} (\bibinfo {year}
  {2019})}\BibitemShut {NoStop}%
\bibitem [{\citenamefont {Rodin}\ \emph {et~al.}(2014)\citenamefont {Rodin},
  \citenamefont {Carvalho},\ and\ \citenamefont {Neto}}]{rodin2014excitons}%
  \BibitemOpen
  \bibfield  {author} {\bibinfo {author} {\bibfnamefont {A.}~\bibnamefont
  {Rodin}}, \bibinfo {author} {\bibfnamefont {A.}~\bibnamefont {Carvalho}},\
  and\ \bibinfo {author} {\bibfnamefont {A.~C.}\ \bibnamefont {Neto}},\
  }\href@noop {} {\bibfield  {journal} {\bibinfo  {journal} {Physical Review
  B}\ }\textbf {\bibinfo {volume} {90}},\ \bibinfo {pages} {075429} (\bibinfo
  {year} {2014})}\BibitemShut {NoStop}%
\bibitem [{\citenamefont {Faria~Junior}\ \emph {et~al.}(2019)\citenamefont
  {Faria~Junior}, \citenamefont {Kurpas}, \citenamefont {Gmitra},\ and\
  \citenamefont {Fabian}}]{PhysRevB.100.115203}%
  \BibitemOpen
  \bibfield  {author} {\bibinfo {author} {\bibfnamefont {P.~E.}\ \bibnamefont
  {Faria~Junior}}, \bibinfo {author} {\bibfnamefont {M.}~\bibnamefont
  {Kurpas}}, \bibinfo {author} {\bibfnamefont {M.}~\bibnamefont {Gmitra}},\
  and\ \bibinfo {author} {\bibfnamefont {J.}~\bibnamefont {Fabian}},\ }\href
  {https://doi.org/10.1103/PhysRevB.100.115203} {\bibfield  {journal} {\bibinfo
   {journal} {Physical Review B}\ }\textbf {\bibinfo {volume} {100}},\ \bibinfo
  {pages} {115203} (\bibinfo {year} {2019})}\BibitemShut {NoStop}%
\end{thebibliography}%

\appendix

\renewcommand\appendixname{Supplementary Note}

\begin{widetext}
\section{The effective exciton Hamiltonian}
The matrix elements of the Hamiltonian (3) between two intralayer exciton basis states are
\begin{equation}\label{eq:Xmelem}
    \braoket{{\rm X_{\lambda',n'}(\QQ')}}{H_{\rm m}}{{\rm X_{\lambda,n}(\QQ)}} = \delta_{\lambda',\lambda}\int d^2r\,\int d^2\rho\,\frac{e^{i(\QQ-\QQ')\cdot\rr}}{\mathcal{S}}X_{n'}^*(\rho)X_{n}(\rho)\Big[ \varepsilon_c(\rr_e[\rrho,\rr]) - \varepsilon_v(\rr_h[\rrho,\rr]) \Big],
\end{equation}
where the electron- and hole position vectors are
\begin{equation}
\begin{split}
    \rr_e[\rrho,\rr]=\unitx\left(\frac{m_x^v}{m_x^c+m_x^v}\rrho\cdot\unitx + \rr\cdot\unitx \right) + \unity\left(\frac{m_y^v}{m_y^c+m_y^v}\rrho\cdot\unity + \rr\cdot\unity \right),\\
    \rr_h[\rrho,\rr]=\unitx\left(-\frac{m_x^c}{m_x^c+m_x^v}\rrho\cdot\unitx + \rr\cdot\unitx \right) + \unity\left(-\frac{m_y^c}{m_y^c+m_y^v}\rrho\cdot\unity + \rr\cdot\unity \right),
\end{split}
\end{equation}
and the $\Gamma$-point electronic energies can be expressed as ($\alpha=c,\,v$)
\begin{equation}
    \varepsilon_\alpha(\rr) = \varepsilon_\alpha^{(0)} + \delta \varepsilon_\alpha(\rr).
\end{equation}
Here, $\varepsilon_\alpha^{(0)}$ is the $\alpha$-band edge energy in a phosphorene monolayer, and $\delta\varepsilon_\alpha(\rr)$ its spatial energy variation due to the presence of the other (twisted) layer.

Exploiting the scale separation between the exciton Bohr radius and the moir\'e length scale ($|a_\ell^{\rm M}|\gg a_{\rm X}$) we approximate
\begin{equation}
    \delta\varepsilon_c (\rr_e[\rrho,\rr]) \approx \delta\varepsilon_c(\rr),\quad \delta\varepsilon_v (\rr_h[\rrho,\rr]) \approx \delta\varepsilon_h(\rr).
\end{equation}
Substituting into \eqref{eq:Xmelem} gives
\begin{equation}\label{eq:Xmelem2}
\begin{split}
    \braoket{{\rm X_{\lambda',n'}(\QQ')}}{H_{\rm m}}{{\rm X_{\lambda,n}(\QQ)}} \approx&\, \delta_{\lambda',\lambda}\int d^2\rho\,X_{n'}^*(\rho)X_{n}(\rho)\int d^2r\,\frac{e^{i(\QQ-\QQ')\cdot\rr}}{\mathcal{S}}\Big[ \varepsilon_c(\rr) - \varepsilon_v(\rr) \Big]\\
    =&\,\delta_{\lambda',\lambda}\delta_{n',n}\Big[\varepsilon_c(\QQ'-\QQ) - \varepsilon_v(\QQ'-\QQ) \Big],
\end{split}
\end{equation}
where $\varepsilon_\alpha(\QQ)$ is the Fourier transform of the position-dependent energy $\varepsilon_\alpha(\rr)$. Analogously, we obtain the matrix element
\begin{equation}
\begin{split}
    \braoket{{\rm IX}_{\lambda',n'}^{\bar{\lambda}'}(\QQ')}{H_{\rm m}}{{\rm X}_{\lambda, n}(\QQ)}= \Bigg[& \int d^2\rho\,Y_{n'}^*(\rrho) X_{n}(\rrho)
    \Bigg]\Big[\delta_{\lambda,\lambda'}\Big(\delta_{\lambda,b}T_c(\QQ'-\QQ) + \delta_{\lambda,t}T_c^*(\QQ'-\QQ)\Big)\\
    &- \delta_{\lambda,\bar{\lambda}'}\Big(\delta_{\lambda,t}T_v(\QQ'-\QQ) + \delta_{\lambda,b}T_v^*(\QQ'-\QQ) \Big) \Big],
\end{split}
\end{equation}
with $T_\alpha(\QQ)$ the Fourier transform of $T_\alpha(\rr)$. As mentioned in the main text, we shall be concerned only with $s$ type excitons, whose RM wavefunctions can always be chosen to be real valued. This allows us to define the renormalized tunneling functions
\begin{equation}\label{RenormT}
    \tilde{T}_{\alpha}(\rr) \equiv T_{\alpha}(\rr)\int d^2\rho\,Y_{1s}(\rrho) X_{1s}(\rrho).
\end{equation}
With this we have
\begin{equation}\label{eq:IXmelem}
\begin{split}
    \braoket{{\rm IX}_{\lambda',n'}^{\bar{\lambda}'}(\QQ')}{H_{\rm m}}{{\rm X}_{\lambda, n}(\QQ)}= \delta_{\lambda,\lambda'}\Big(\delta_{\lambda,b}\tilde{T}_c(\QQ'-\QQ) + \delta_{\lambda,t}\tilde{T}_c(\QQ'-\QQ)\Big) - \delta_{\lambda,\bar{\lambda}'}\Big(\delta_{\lambda,t}\tilde{T}_v(\QQ'-\QQ) + \delta_{\lambda,b}\tilde{T}_v(\QQ'-\QQ) \Big).
\end{split}
\end{equation}
Henceforth, we shall drop the RM wave function index $1s$, for simplicity.

To obtain an effective Hamiltonian for the excitons, we supplement the moir\'e Hamiltonian $H_{\rm m}$ with the excitons' COM kinetic energies $K(\QQ)$ (see Sec.\ \ref{Sec:HamiltonianDiag} below) and COM-position-dependent binding energies $\epsilon_\mu(\rr)$, where $\mu = {\rm X},\,{\rm IX}$. Here, the COM wave vector $\QQ$ is written as a column vector, $\QQ^T$ is its transpose, and $M^{-1}$ is the anisotropic total mass tensor. We now define the four-spinor
\begin{equation}
    \Xcal(\QQ)\equiv \begin{pmatrix}
        \ket{{\rm X}_{b}(\QQ)}\\ \ket{{\rm X}_t(\QQ)} \\ \ket{{\rm IX }_b^t(\QQ)} \\ \ket{{\rm IX }_t^b(\QQ)}
    \end{pmatrix},
\end{equation}
which gives the effective exciton Hamiltonian in the form
\begin{equation}
\begin{split}
    H_{\rm m}(\QQ',\QQ) = \delta_{\QQ',\QQ}\Xcal^\dagger(\QQ) K(\QQ)\Xcal(\QQ) + \int d^2r\,\frac{e^{-i(\QQ'-\QQ)\cdot\rr}}{\Scal} \Xcal^\dagger (\QQ') \mathcal{H}_{\rm m}(\rr) \Xcal(\QQ),
\end{split}
\end{equation}
with the real-space moir\'e Hamiltonian
\begin{equation}\label{eq:Hrealspace}
    \mathcal{H}_{\rm m}(\rr) = \begin{pmatrix}
        \Ecal_{{\rm X}}(\rr) & 0 & \tilde{T}_c(\rr) & -\tilde{T}_v(\rr) \\
        0 & \Ecal_{{\rm X}}(\rr) & -\tilde{T}_v(\rr) & \tilde{T}_c(\rr)\\
        \tilde{T}_c(\rr) & -\tilde{T}_v(\rr) & \Ecal_{{\rm IX}}(\rr) & 0 \\
        -\tilde{T}_v(\rr) & \tilde{T}_c(\rr) & 0 & \Ecal_{{\rm IX}}(\rr)
    \end{pmatrix},
\end{equation}
reported in Eq.\ (6) of the main text. Here, we have defined the exciton potentials
\begin{equation}
    \Ecal_{\mu}(\rr) \equiv \varepsilon_c(\rr) - \varepsilon_v(\rr) - \varepsilon_\mu(\rr) = \Ecal^{(0)} + \delta\varepsilon_c(\rr) - \delta\varepsilon_v(\rr) + \epsilon_\mu(\rr),
\end{equation}
with $\Ecal^{(0)} = \varepsilon_c^{(0)}-\varepsilon_v^{(0)}$ the phosphorene monolayer band gap.

\section{Hamiltonian Fourier components}

The real space Hamiltonian for electrons and holes presented in the main text is composed by two elements: the intralayer moiré potential
\begin{subequations}\label{IntraInt}
\begin{equation}
    \varepsilon_{\alpha}^{\lambda}(\rr) = \varepsilon_{\alpha}^{(0)} + \delta\varepsilon_{\alpha}(\rr),
\end{equation}
\begin{equation}
    \delta\varepsilon_{\alpha}(\rr) = \sum_{n=0}^{6} \varepsilon_{\alpha}^{(n)}e^{-q_{n}[d(\rr)-d_{0}]}\cos(\gG_{n}\cdot\rr),
\end{equation}
\end{subequations}
and the tunneling terms
\begin{equation}\label{TunInt}
    T_{\alpha}(\rr) = \sum_{n=0}^{6} t_{\alpha}^{(n)}e^{-q_{n}[d(\rr)-d_{0}]}\cos(\gG_{n}\cdot\rr).
\end{equation}
Here, we have the constraints $\varepsilon_{\alpha}^{(1)}=\varepsilon_{v}^{(3)}=\varepsilon_{\alpha}^{(5)}=t_{v}^{(0)}=t_{v}^{(5)}=t_{c}^{(1)}=t_{c}^{(5)}=0$, $\varepsilon_{\alpha}^{(3)} = \varepsilon_{\alpha}^{(4)}$ and $t_{\alpha}^{(3)}=t_{\alpha}^{(4)}$, for $\alpha=c,v$. The coefficients of the tunneling and intralayer moiré potentials contain an exponential decay with respect to the interlayer distance $d$, which in turn varies along the material's plane according to the function
\begin{equation}\label{distFit}
    d(\rr) = d_{0} + \sum_{n=1}^{4}\big[ d_{n}^{s}\cos(\gG_{n}\cdot\rr) + d_{n}^{a}\sin(\gG_{n}\cdot\rr)\big].
\end{equation}
Parameters in \eqref{IntraInt}, \eqref{TunInt} and \eqref{distFit} were fitted to DFT calculations in Ref. \onlinecite{MoirePhos.PhysRevB.105.235421} and are shown in Table \ref{tab:FitVal}.

\begin{table}[]
    \centering
    \caption{Parameters of the real space Hamiltonian for the twisted phosphorene bilayer and the interlayer distance $d$. Each Fourier series coefficient was fitted to DFT calculations as $A(d)=A_{0}e^{-q_{n}(d-d_{0})}$, with $d_{0}=3.49$ \AA.}
    \begin{tabular}{c c c | c c c | c c}
    \hline
    \hline
       $A$ & $A_{0}$ [eV] & $q$ [\AA${}^{-1}$]  & $A$ & $A_{0}$ [eV] & $q$ [\AA${}^{-1}$] &  & [\AA] \\
    \hline
       $t_c^{(0)}$ &  0.384 & 0.61 & $\varepsilon_c^{(0)}$ & 0.360 & 0.77 & $d_{1}^{s}$ & -0.016  \\
$t_c^{(2)}$ & -0.185 & 1.37 & $\varepsilon_c^{(2)}$ & -0.094 & 1.17 &  $d_{2}^{s}$ & -0.124 \\
$t_c^{(3)}$ &  0.003 & 2.72 & $\varepsilon_c^{(3)}$ & -0.011 & 1.11 & $d_{3}^{s}$ & 0.088  \\
$t_c^{(6)}$ &  0.013 & 2.50 & $\varepsilon_c^{(6)}$ &  0.058 & 1.49 & $d_{4}^{s}$ & 0.088 \\
$t_v^{(1)}$ &  0.023 & 1.11 & $\varepsilon_v^{(0)}$ & -0.209 & 0.00 & $d_{1}^{a}$ & 0.072  \\
$t_v^{(2)}$ &  0.266 & 1.34 & $\varepsilon_v^{(2)}$ & -0.068 & 0.89 & $d_{2}^{a}$ & 0.150 \\
$t_v^{(3)}$ & -0.010 & 2.90 & $\varepsilon_v^{(6)}$ &  0.136 & 2.08 & $d_{3}^{a}$ & 0.021 \\
$t_v^{(6)}$ & -0.022 & 2.25 & \, & & & $d_{4}^{a}$ & 0.062 \\
    \hline
    \hline
    \end{tabular}
    \label{tab:FitVal}
\end{table}

Field operators can be expanded in terms of the eigenstates around the $\Gamma$ point
\begin{equation}\label{FieldExp}
    \varphi_{\alpha\lambda}(\rr) = \sum_{\kk} \frac{e^{i\kk\cdot\rr}}{\sqrt{\mathcal{S}}}\, c_{\alpha\lambda}(\kk),
\end{equation}
leading to the following Hamiltonian:
\begin{equation}
\begin{split}
    H_{\rm m} = \sum_{\alpha,\lambda}\sum_{\kk,\kk'}\varepsilon_{\alpha}^{\lambda}(\kk-\kk')c_{\alpha\lambda}^{\dagger}(\kk)c_{\alpha\lambda}(\kk') + \sum_{\alpha}\sum_{\kk,\kk'}\Big[ T_{\alpha}(\kk-\kk')c_{\alpha b}^{\dagger}(\kk)c_{\alpha t}(\kk') + \text{H.c.} \Big],
\end{split}
\end{equation}
where we apply a second order approximation over the exponential terms in \eqref{TunInt}, obtaining the following tunneling matrix elements
\begin{equation}\label{InterME}
\begin{split}
    T_{\alpha}(\kk-\kk') \equiv&\,  \int\frac{d^{2}r}{\mathcal{S}} \, T_{\alpha}(\rr)e^{-i(\kk-\kk')\cdot\rr}\\
    \approx&\, \sum_{n=0}^{6}\frac{t_{\alpha}^{(n)}}{2}\Bigg[ \delta_{\kk,\kk'+\gG_{n}} + \delta_{\kk,\kk'-\gG_{n}}- \frac{q_{n}}{2}\sum_{m=1}^{4}(d_{m}^{s}+id_{m}^{a})(\delta_{\kk,\kk'+\gG_{n}+\gG_{m}}+\delta_{\kk,\kk'-\gG_{n}+\gG_{m}})\\
    & \qquad + \frac{q_{n}}{2}\sum_{m=1}^{4}(-d_{m}^{s}+id_{m}^{a})(\delta_{\kk,\kk'+\gG_{n}-\gG_{m}}+\delta_{\kk,\kk'-\gG_{n}-\gG_{m}})\\
    & \qquad+ \frac{q_{n}^{2}}{8}\sum_{m=1}^{4}\sum_{\ell=1}^{4}\Big\{ (d_{m}^{s}d_{\ell}^{s} + 2id_{m}^{s}d_{\ell}^{a} - d_{m}^{a}d_{\ell}^{a})(\delta_{\kk,\kk'+\gG_{n}+\gG_{m}+\gG_{\ell}} + \delta_{\kk,\kk'-\gG_{n}+\gG_{m}+\gG_{\ell}})\\
    & \qquad + (d_{m}^{s}d_{\ell}^{s} - 2id_{m}^{s}d_{\ell}^{a} + d_{m}^{a}d_{\ell}^{a})(\delta_{\kk,\kk'+\gG_{n}+\gG_{m}-\gG_{\ell}} + \delta_{\kk,\kk'-\gG_{n}+\gG_{m}-\gG_{\ell}})\\
    &\qquad + (d_{m}^{s}d_{\ell}^{s} + 2id_{m}^{s}d_{\ell}^{a} + d_{m}^{a}d_{\ell}^{a})(\delta_{\kk,\kk'+\gG_{n}-\gG_{m}+\gG_{\ell}} + \delta_{\kk,\kk'-\gG_{n}-\gG_{m}+\gG_{\ell}})\\
    &\qquad +(d_{m}^{s}d_{\ell}^{s} - 2id_{m}^{s}d_{\ell}^{a} - d_{m}^{a}d_{\ell}^{a})(\delta_{\kk,\kk'+\gG_{n}-\gG_{m}-\gG_{\ell}} + \delta_{\kk,\kk'-\gG_{n}-\gG_{m}-\gG_{\ell}})\Big\}\Bigg].
\end{split}
\end{equation}
The intralayer matrix elements are
\begin{equation}
\begin{split}
    \varepsilon_{\alpha}^{\lambda}(\kk-\kk') \equiv&\, \int \frac{d^{2}r}{\mathcal{S}}\, \varepsilon_{\alpha}^{\lambda}(\rr)e^{-i(\kk-\kk')\cdot\rr}\\
    =&\, \varepsilon_{\alpha}^{\lambda(0)}(\kk)\delta_{\kk,\kk'} + \delta\varepsilon_{\alpha}(\kk'-\kk),
\end{split}
\end{equation}
where
\begin{equation}
    \varepsilon_\alpha^{t/b(0)}(\kk) \equiv \varepsilon_\alpha^{(0)} + \frac{\hbar^{2}|(\mathcal{R}_{\mp\theta/2}\kk)\cdot\hat{\mathbf{x}}|^{2}}{2m_{\alpha,x}} + \frac{\hbar^{2}|(\mathcal{R}_{\mp\theta/2}\kk)\cdot\hat{\mathbf{y}}|^{2}}{2m_{\alpha,y}},
\end{equation}
and the terms $\delta\varepsilon_{\alpha}(\kk-\kk')$ have the same structure as \eqref{InterME}.

\section{Hamiltonian diagonalization}\label{Sec:HamiltonianDiag}

In order to diagonalize the exciton Hamiltonian, we evaluate the matrix elements between the different states. Consider the exciton states in their Fourier representation
\begin{subequations}\label{eq:Xwfk}
\begin{equation}
    \ket{\text{X}_{\lambda}(\QQ)} = \, \sum_{\qq}\frac{\tilde{X}(\qq)}{\sqrt{\mathcal{S}}}  c_{c\lambda}^{\dagger}\big( \kk_{e}[\qq,\QQ]\big)c_{v\lambda}\big( \kk_{h}[\qq,\QQ]\big)\ket{\Omega},
\end{equation}
\begin{equation}            
    \ket{\text{IX}_{\lambda}^{\bar{\lambda}}(\QQ)} = \, \sum_{\qq}\frac{\tilde{Y}(\qq)}{\sqrt{\mathcal{S}}}  c_{c\Bar{\lambda}}^{\dagger}\big( \kk_{e}[\qq,\QQ]\big)c_{v\lambda}\big( \kk_{h}[\qq,\QQ]\big)\ket{\Omega},
\end{equation}
\end{subequations}
with
\begin{subequations}\label{kQwavevec}
\begin{equation}
    \kk_{e}[\qq,\QQ] = \bigg( q_{x} + \frac{m_{c,x}}{M_{x}}Q_{x}, q_{y} + \frac{m_{c,y}}{M_{y}}Q_{y} \bigg),
\end{equation}
\begin{equation}
    \kk_{h}[\qq,\QQ] =\bigg(q_{x} - \frac{m_{v,x}}{M_{x}}Q_{x}, q_{y} - \frac{m_{v,y}}{M_{y}}Q_{y}\bigg),
\end{equation}
\end{subequations}
and $\tilde{X}(\qq)$, $\tilde{Y}(\qq)$ are the Fourier transforms of the corresponding RM wave functions. The matrix element between a X and an IX state is
\begin{equation}
\begin{split}
    \braoket{\text{X}_{b}(\QQ)}{H_{\rm m}}{\text{IX}_{b}^{t}(\QQ')} =& \, \frac{1}{\mathcal{S}}\sum_{\kk,\kk'}\sum_{\qq,\qq'} \tilde{X}^{*}(\qq)\tilde{Y}(\qq')T_{c}(\kk'-\kk)\delta_{\kk',\kk_{e}'}\delta_{\kk_{e},\kk}\delta_{\kk_{h},\kk_{h}'}\\
    =&\, T_{c}(\QQ-\QQ') \int d^{2}r\, \text{exp}\Bigg[ i\sum_{\zeta=x,y} \frac{m_{v,\zeta}}{M_{\zeta}}(Q_{\zeta} - Q_{\zeta}')\zeta \Bigg]X^{*}(\rr)Y(\rr).
\end{split}
\end{equation}
Taking into account the separation of scales between the exciton spacial extension and the mSC for small twist angles, it is possible to apply the envelope approximation in the exponential term, obtaining
\begin{equation}
    \braoket{\text{X}_{b}(\QQ)}{H_{\rm m}}{\text{IX}_{b}^{t}(\QQ')} \approx T_{c}(\QQ- \QQ') \int d^{2}r\, X^{*}(\rr)Y(\rr) =\, \tilde{T}_{c}(\QQ-\QQ').
\end{equation}
Similarly, the rest of the matrix elements between X and IX excitons are
\begin{subequations}\label{ExcTElem2}
\begin{equation}
    \braoket{\text{X}_{t}(\QQ)}{H_{\rm m}}{\text{IX}_{t}^{b}(\QQ')} \approx\, \tilde{T}_{c} (\QQ-\QQ'),
\end{equation}
\begin{equation}
    \braoket{\text{X}_{t}(\QQ)}{H_{\rm m}}{\text{IX}_{b}^{t}(\QQ')} = \braoket{\text{X}_{b}(\QQ)}{H_{\rm m}}{\text{IX}_{t}^{b}(\QQ')} \approx\, -\tilde{T}_{v}(\QQ-\QQ').
\end{equation}
\end{subequations}
Considering the same approximations, the matrix elements between X excitons of the same species are
\begin{equation}\label{HamElemX}
    \mathcal{E}_{\rm X_{\lambda}}[\QQ,\QQ'] \equiv \braoket{\text{X}_{\lambda}(\QQ)}{H_{\rm m}}{\text{X}_{\lambda}(\QQ')} = \big[\varepsilon_{c}^{\lambda\,(0)}- \varepsilon_{v}^{\lambda\,(0)}\big]\delta_{\QQ\QQ'} + \delta\varepsilon_{c}(\QQ-\QQ')  - \delta\varepsilon_{v}(\QQ-\QQ'),
\end{equation}
where $\varepsilon_{\alpha}^{\lambda\,(0)}$ is the quasiparticle dispersion energy for band $\alpha=c,v$ in layer $\lambda=b,t$ around the $\Gamma$ point. The excitation energy $\varepsilon_{c}^{\lambda\,(0)} - \varepsilon_{v}^{\lambda\,(0)}$ can be written in terms of the CoM and relative motion energies as
\begin{equation}
    \varepsilon_{c}^{\lambda\,(0)} - \varepsilon_{v}^{\lambda\,(0)} = \mathcal{E}_{\rm X_{\lambda}}^{(0)}(\QQ) + \epsilon_{\rm X},
\end{equation}
with $\epsilon_{X}$ the intralayer binding energy obtained considering the inverse reduced mass tensor
\begin{equation}\label{ReducedMass0}
    \mu_0^{-1} = \begin{pmatrix}
        \frac{M_{x}}{m_x^cm_x^v} & 0 \\ 0 & \frac{M_{y}}{m_y^cm_y^v}
    \end{pmatrix},
\end{equation}
and the CoM dispersion
\begin{equation}\label{eq:Xdisp0}
    \mathcal{E}_{\rm X_{b/t}}^{(0)}(\QQ) \equiv \, \delta E_{\rm sc} + \frac{\hbar^{2}|(\mathcal{R}_{\pm\theta/2}\QQ)\cdot\hat{\mathbf{x}}|^{2}}{2M_{x}} + \frac{\hbar^{2}|(\mathcal{R}_{\pm\theta/2}\QQ)\cdot\hat{\mathbf{y}}|^{2}}{2M_{y}},
\end{equation}
with $M_x=m_x^c+m_x^v$, $M_y=m_y^c+m_y^v$. The $\mathcal{R}_{\varphi}$ operator represents a rotation by an angle $\varphi$ about the $\hat{\bvec{\mathrm{z}}}$ axis, and the reference energy $\delta E_{\rm sc} = 1.31$ eV is a scissor-correction to reproduce the band gap across the mSC. Equation \eqref{eq:Xdisp0} can be rewritten in matrix form as
\begin{equation}\label{eq:Xdisp1}
    \mathcal{E}_{\rm X_{b/t}}^{(0)}(\QQ) = \delta E_{\rm sc} + \frac{\hbar^2}{2}\QQ^T \mathcal{R}_{\mp\theta/2} M_0^{-1} \mathcal{R}_{\pm\theta/2} \QQ,
\end{equation}
with the inverse total mass tensor
\begin{equation}
    M_0^{-1} = \begin{pmatrix}
        M_x^{-1} & 0 \\ 0 & M_y^{-1}
    \end{pmatrix}.
\end{equation}
We may define the rotated inverse total mass tensor
\begin{equation}
    M_{\pm\theta/2}^{-1} \equiv \mathcal{R}_{\mp\theta/2} M_0^{-1} \mathcal{R}_{\pm\theta/2} = M_0^{-1} \mp \theta (M_x^{-1} - M_y^{-1}) \begin{pmatrix}
        0 & 1 \\ 1 & 0 
    \end{pmatrix} + \mathcal{O}\{\theta^2\}.
\end{equation}
The maximal error incurred by approximating $M_{\pm\theta/2}^{-1}\approx M_0^{-1}$ occurs at the maximum twist angle for which our model is valid, $\theta = 6^\circ$, and can be estimated by computing the principal axes of $M_{\pm\theta/2}^{-1}$, $\hat{\mathbf{v}}_1$ and $\hat{\mathbf{v}}_2$, and comparing them with those of $M_0^{-1}$, $\hat{\mathbf{x}}$ and $\hat{\mathbf{y}}$. This yields
\begin{equation}
    |\hat{\mathbf{v}}_1 - \hat{\mathbf{x}}| = |\hat{\mathbf{v}}_2 - \hat{\mathbf{y}}| = 0.17.
\end{equation}
This error in principal axes propagates to the CoM momenta, which appear squared in the CoM dispersion, thus resulting in a total error of $3\%$ in the CoM dispersion, indicating that $M_{\pm\theta/2}^{-1}\approx M_0^{-1}$ is a good approximation. Therefore, we approximate
\begin{equation}\label{eq:XdispFinal}
    \mathcal{E}_{\rm X_{b/t}}^{(0)}(\QQ) \approx \delta E_{\rm sc} + \frac{\hbar^2}{2}\QQ^T  M_0^{-1}  \QQ.
\end{equation}

The matrix elements between IX excitons have the same structure \eqref{HamElemX}, with the excitation energy for electron and hole in different layers given by
\begin{equation}
\begin{split}
    \varepsilon_{c}^{t\,(0)} - \varepsilon_{v}^{b\,(0)} =&\, \delta E_{\rm sc} + \frac{\hbar^{2}|(\mathcal{R}_{-\theta/2}\kk_{e})\cdot\hat{\mathbf{x}}|^{2}}{2m_{c,x}} + \frac{\hbar^{2}|(\mathcal{R}_{-\theta/2}\kk_{e})\cdot\hat{\mathbf{y}}|^{2}}{2m_{c,y}} - \frac{\hbar^{2}|(\mathcal{R}_{+\theta/2}\kk_{h})\cdot\hat{\mathbf{x}}|^{2}}{2m_{v,x}} - \frac{\hbar^{2}|(\mathcal{R}_{+\theta/2}\kk_{h})\cdot\hat{\mathbf{y}}|^{2}}{2m_{v,y}}\\
    =&\, \delta E_{\rm sc} + \frac{\hbar^{2}Q_{x}^{2}}{2M_{x}} + \frac{\hbar^{2}Q_{y}^{2}}{2M_{y}} + \frac{\hbar^{2}q_{x}^{2}}{2\mu_{x}} + \frac{\hbar^{2}q_{y}^{2}}{2\mu_{y}} + \frac{\hbar^{2}\theta}{2}\bigg[ \bigg( \frac{1}{m_{c,x}} - \frac{1}{m_{c,y}} + \frac{1}{m_{v,y}} - \frac{1}{m_{v,x}} \bigg) q_{x}q_{y}\\
    & +\bigg\{ \mu_{y} \bigg( \frac{1}{m_{c,x}m_{v,y}} + \frac{1}{m_{c,y}m_{v,x}} \bigg) - \frac{2}{M_{y}} \bigg\} q_{x}Q_{y} + \bigg\{ \frac{2}{M_{x}} - \mu_{x}\bigg( \frac{1}{m_{c,y}m_{v,x}} + \frac{1}{m_{c,x}m_{v,y}} \bigg) \bigg\} q_{y}Q_{x} \\
    & + \frac{m_{c,y} - m_{c,x} + m_{v,x} - m_{v,y}}{M_{x}M_{y}}\,Q_{x}Q_{y} \bigg] + \mathcal{O}\{\theta^2\},
\end{split}
\end{equation}
where the electron and hole wave vectors, $\kk_{e}$ and $\kk_{h}$, have been replaced using \eqref{kQwavevec}. The above expression can be written in matrix form as
\begin{equation}\label{eq:dispMform}
    \varepsilon_{c}^{t\,(0)} - \varepsilon_{v}^{b\,(0)} = \delta E_{\rm sc} + \frac{\hbar^2}{2}\left( \QQ^T M^{-1} \QQ + \qq^T \mu^{-1} \qq + \QQ^T m^{-1} \qq + \qq^T \left[m^{-1}\right]^T \QQ \right),
\end{equation}
with the definitions
\begin{subequations}
\begin{equation}
    M^{-1} = M_0^{-1} + \theta\frac{m_{c,y} - m_{c,x} + m_{v,x} - m_{v,y}}{2M_{x}M_{y}}\begin{pmatrix}
        0 &1 \\ 1 & 0
    \end{pmatrix},
\end{equation}
\begin{equation}
    \mu^{-1} = \mu_0^{-1} + \frac{\theta}{2} \bigg( \frac{1}{m_{c,x}} - \frac{1}{m_{c,y}} + \frac{1}{m_{v,y}} - \frac{1}{m_{v,x}} \bigg)\begin{pmatrix}
        0 &1 \\ 1 & 0
    \end{pmatrix},
\end{equation}
\begin{equation}
    m^{-1} = \frac{\theta}{2}\begin{pmatrix}
        0 & \frac{2}{M_{x}} - \mu_{x}\bigg[ \frac{1}{m_{c,y}m_{v,x}} + \frac{1}{m_{c,x}m_{v,y}} \bigg] \\
        \mu_{y} \bigg[ \frac{1}{m_{c,x}m_{v,y}} + \frac{1}{m_{c,y}m_{v,x}} \bigg] - \frac{2}{M_{y}} & 0
    \end{pmatrix},
\end{equation}
\end{subequations}
We may now estimate the error incurred by dropping the terms proportional to $\theta$ in the same manner as for the intralayer exciton CoM dispersions, giving a total error of $1.4\%$. Moreover, the same procedure yields an error below $0.1\%$ for the RM dispersion. For simplicity, we henceforth approximate $M^{-1}\approx M_0^{-1}$ and $\mu^{-1}\approx \mu_0^{-1}$ when describing the IX energies. 

Finally, the last two terms in Eq.\ \eqref{eq:dispMform} introduce a perturbation of maximal magnitude $\sim 0.07q_*Q_{*} \hbar^2/m_0=64\,{\rm meV}$ at $\theta=6^\circ$ for the lowest exciton states, where $Q_*\sim|\mathbf{g}_1|=0.2\,{\rm \AA}^{-1}$, and $q_*\sim 2\pi/a_{{\rm B},y}=0.6\,{\rm \AA}^{-1}$, with $a_{{\rm B},y}=10\,{\rm \AA}$ the typical maximal exciton Bohr length in hBN-encapsulated phosphorene, as computed using our methods. Within the range of validity of our model, this perturbation is small compared with both the electron-hole interactions, characterized by exciton binding energies of order $100\,{\rm meV}$ [see main text Fig.\ 2(a)], as well as the moir\'e potential, with a total amplitude of approximately $600\,{\rm meV}$ [see main text Fig.\ 2(c)], and we neglect it as a first approximation.

Following the approximations described above, we obtain the IX energy in the form
\begin{equation}
    \mathcal{E}_{\rm IX}[\QQ,\QQ']\equiv\braoket{\text{IX}_{\lambda}^{\bar{\lambda}}(\QQ)}{H_{\rm m}}{\text{IX}_{\lambda}^{\bar{\lambda}}(\QQ')} \approx\, \mathcal{E}_{\rm IX}^{(0)}(\QQ)\delta_{\QQ\QQ'} + \delta\varepsilon_{c,\QQ\QQ'}  - \delta\varepsilon_{v,\QQ\QQ'} + \epsilon_{\rm IX},
\end{equation}
with
\begin{equation}
    \mathcal{E}_{\rm IX}^{(0)}(\QQ) =\, \delta E_{\rm sc} + \frac{\hbar^2}{2}\QQ^T M_0^{-1} \QQ,
\end{equation}
and $\epsilon_{\rm IX}$ the binding energy obtained using the inverse reduced mass tensor $\mu_0^{-1}$. Lastly, as the interlayer distance vaires across the mSC, so do the eigenfuctions $X(\rr)$ and $Y(\rr)$, including the overlap integral in Eq. \eqref{RenormT}. The calculation this overlap along the supercell indicates that it varies between $0.8923$ and $0.9082$, which generates negligible energy fluctuations when multiplying by $T_{\alpha}$. Instead, we consider the average value $0.9014$ with respect to 13 interlayer distances representative of the mSC.

\section{Electron-hole relative motion problem}\label{sec:RelativeMotion}

The electron-hole system is described through the anisotropic Wannier equation
\begin{equation}\label{AnWann}
    \bigg[ -\frac{\hbar^{2}}{2\mu_{x}}\frac{\partial^{2}}{\partial x^{2}} -\frac{\hbar^{2}}{2\mu_{y}}\frac{\partial^{2}}{\partial y^{2}} + U_{\lambda}^{\lambda'}(\bvec{\rho}) \bigg] \psi(\bvec{\rho}) = \epsilon \psi(\bvec{\rho}),
\end{equation}
where, as described in Sec. \ref{Sec:HamiltonianDiag}, we consider the reduced mass tensor $\mu_{0}$ (see Eq. \eqref{ReducedMass0}) for both X and IX, $\bvec{\rho} = \rr_{e} - \rr_{h}$ is the relative position vector, and $U_\lambda^{\lambda'}$ is the screened electrostatic interaction between a hole in layer $\lambda$ and an electron in layer $\lambda'$. Naturally, the electron-hole interaction in an anisotropic material is anisotropic as well, and depends on the dielectric tensor $\epsilon_{ij}$ of the medium surrounding the bilayer, and on the in-plane electrical polarizability of the material along the $x$ and $y$ directions, denoted by $\kappa_{x}$ and $\kappa_{y}$, respectively. Accroding to Ref.\ \cite{rodin2014excitons}, $\kappa_{x} = 3.97$ \AA$\,$ and $\kappa_{y} = 4.20$ \AA$\,$ for the phosphorene monolayer. These values differ by only $6\%$, by contrast to the $77\%$ difference between the reduced masses $\mu_x=0.660m_0$ and $\mu_y=0.153m_0$, indicating that the anisotropy of \eqref{AnWann} is dominated by the kinetic energy term. Therefore, we simplify the problem by using an isotropic polarizability $\kappa = (\kappa_{x} + \kappa_{y})/2$, without changing the binding energies significantly. Furthermore, the bilayer will be considered to be immersed in an anisotropic medium with dielectric tensor $\epsilon = \text{diag}(\epsilon_{\parallel}, \epsilon_{\parallel}, \epsilon_{\perp})$, where $\epsilon_{\parallel}$ is the permitivity in the $xy$ plane, and $\varepsilon_{\perp}$ along $z$.

The interaction $U_{\lambda}^{\lambda'}$ between charge carriers does not have a closed analytic real-space form, but can be expressed in terms of its Fourier components as \cite{danovich2018localized,viner2021excited}:
\begin{subequations}\label{KeldPot}
\begin{equation}
    U_{\lambda}^{\lambda}(\qq) = -\frac{2\pi}{\Tilde{\varepsilon}q}\frac{1 + r_{*}q - r_{*}q e^{-2q\Tilde{d}}}{(1 + r_{*}q)^{2} - r_{*}^{2}q^{2}e^{-2q\Tilde{d}}},
\end{equation}
\begin{equation}
    U_{\lambda}^{\bar{\lambda}}(\qq) = -\frac{2\pi}{\Tilde{\varepsilon}q}\frac{e^{-q\Tilde{d}}}{(1 + r_{*}q)^{2} - r_{*}^{2}q^{2}e^{-2q\Tilde{d}}},
\end{equation}
\end{subequations}
where $\qq$ is the wave vector, $\Tilde{\varepsilon}=\sqrt{\varepsilon_{\parallel}\varepsilon_{\perp}}$ is the effective dielectric constant for the medium, $r_{*}=2\pi\kappa/\Tilde{\varepsilon}$ is the screening length, and $\Tilde{d}=d \sqrt{\varepsilon_{\parallel}/\varepsilon_{\perp}}$ is the renormalized interlayer distance.

In order to solve the Wannier equation with potentials \eqref{KeldPot}, we employ a method introduced by Griffin and Wheeler \cite{griffin1957collective} which consists on writting the eigenfunctions of the Hamiltonian in terms of an appropriate finite basis that allows the problem to be diagonalized numerically. In analogy with the analytical solution of the 2D hydrogen atom, we write the eigenfunctions in terms of the basis
\begin{equation}
    \phi_{jm}(\rho,\varphi) = \rho^{|m|}e^{-\beta_{j}\rho}\frac{e^{im\varphi}}{\sqrt{2\pi}}
\end{equation}
where $\bvec{\rho} = (\rho,\varphi)$ is written in polar coordinates and $m\in\mathbb{Z}$ is the magnetic quantum number. This functions capture the expected asymptotic behavior for $\rho\rightarrow 0$ and $\rho\rightarrow \infty$, and each one is characterized by a decay factor $\beta_{j}$ that defines the length scale. The set of values for $\beta_{j}$ is chosen in such a way that the possible values of the exciton Bohr radius are covered, which was achieved through the logarithmic distribution $\beta_{j} = \beta_{1}e^{\eta(j-1)}$, where $\eta=(N-1)^{-1}\ln(\beta_{N}/\beta_{1})$ and $\beta_{N}^{-1}<r_{*}\ll\beta_{1}^{-1}$. Note that, since we have an anisotropic problem, $m$ is no longer a good quantum number. In fact, since the Hamiltonian has a rotational symmetry $C_{2}$, we have a coupling between two functions with quantum numbers $m$ and $m'$ whenever $(m-m')\mod{2}=0$ is satisfied. The appropiate quantum number is then $\Bar{m}=m\mod{2}=0,1$. The approximate solution is written as
\begin{equation}\label{EigExpand}
    \psi_{\Bar{m}} (\rho,\varphi) = \sum_{m}\sum_{j=1}^{N} a_{jm}\phi_{jm}(\rho,\varphi),
\end{equation}
where the sum is restricted to the $m$ values that satisfy $\Bar{m}=m\mod{2}$. Substituting \eqref{EigExpand} in \eqref{AnWann}, we get the generalized eigenvalue problem
\begin{equation}\label{GenEig}
    \big[\bvec{H}_{\Bar{m}} - E \bvec{S}_{\Bar{m}}\big]\bvec{A}_{\Bar{m}} = \bvec{0},
\end{equation}
where $\bvec{H}_{\Bar{m}}$ is the kernel Hamiltonian. The matrix elements are
\begin{equation}
    H_{jj'}^{mm'} = K_{x,jj'}^{mm'} + K_{y,jj'}^{mm'} + U_{jj'}^{mm'},
\end{equation}
with
\begin{subequations}
\begin{equation}
\begin{split}
    K_{x,jj'}^{mm'} =& -\frac{\hbar^{2}}{2\mu_{x}}\int d^{2}\rho\, \phi_{jm}^{*}(\bvec{\rho})\frac{\partial^{2}\phi_{j'm'}(\bvec{\rho})}{\partial x^{2}}\\
    =& -\frac{\hbar^{2}}{2\mu_{x}}\bigg[ \bigg\{ \frac{1}{2}(m^{2}-|m|)\delta_{|m-m'|,2} + \frac{1}{4}m(|m|-1)(\delta_{m',m+2}-\delta_{m',m-2})\bigg\}\frac{\Gamma(|m|+|m'|)}{(\beta_{j} + \beta_{j'})^{|m|+|m'|}}\\
    &\qquad\qquad +\bigg\{ -\frac{1}{2}(2|m|+1)\delta_{m,m'} - \frac{1}{4}(2|m|-1)\delta_{|m-m'|,2} + \frac{1}{2}m(\delta_{m',m-2}-\delta_{m',m+2})\bigg\}\frac{\beta_{j}\Gamma(|m|+|m'|+1)}{(\beta_{j} + \beta_{j'})^{|m|+|m'|+1}}\\
    &\qquad\qquad +\bigg\{ \frac{1}{2}\delta_{m,m'}+\frac{1}{4}\delta_{|m-m'|,2}\bigg\}\frac{\beta_{j}^{2}\Gamma(|m|+|m'|+2)}{(\beta_{j} + \beta_{j'})^{|m|+|m'|+2}}\bigg],
\end{split}
\end{equation}
\begin{equation}
\begin{split}
    K_{y,jj'}^{mm'} =& -\frac{\hbar^{2}}{2\mu_{y}}\int d^{2}\rho\, \phi_{jm}^{*}(\bvec{\rho})\frac{\partial^{2}\phi_{j'm'}(\bvec{\rho})}{\partial y^{2}}\\
    =& -\frac{\hbar^{2}}{2\mu_{y}}\bigg[ \bigg\{ -\frac{1}{2}(m^{2}-|m|)\delta_{|m-m'|,2} - \frac{1}{4}m(|m|-1)(\delta_{m',m+2}-\delta_{m',m-2})\bigg\}\frac{\Gamma(|m|+|m'|)}{(\beta_{j} + \beta_{j'})^{|m|+|m'|}}\\
    &\qquad\qquad +\bigg\{ -\frac{1}{2}(2|m|+1)\delta_{m,m'} + \frac{1}{4}(2|m|-1)\delta_{|m-m'|,2} - \frac{1}{2}m(\delta_{m',m-2}-\delta_{m',m+2})\bigg\}\frac{\beta_{j}\Gamma(|m|+|m'|+1)}{(\beta_{j} + \beta_{j'})^{|m|+|m'|+1}}\\
    &\qquad\qquad +\bigg\{ \frac{1}{2}\delta_{m,m'} - \frac{1}{4}\delta_{|m-m'|,2}\bigg\}\frac{\beta_{j}^{2}\Gamma(|m|+|m'|+2)}{(\beta_{j} + \beta_{j'})^{|m|+|m'|+2}}\bigg],
\end{split}
\end{equation}
\end{subequations}
where $\Gamma(x)$ is the gamma function. The potential energy matrix elements are
\begin{equation}\label{IntKeld}
\begin{split}
    U_{jj'}^{mm'} =& \int d^{2}\rho\, \phi_{jm}^{*}(\bvec{\rho})U(\rho)\phi_{j'm'}(\bvec{\rho}) = \delta_{m,m'}\int_{0}^{\infty} d\rho \, \rho f_{jj'}(\rho)U(\rho)\\
    =&\, 2\pi\delta_{m,m'}\int d^{2}\rho \, f_{jj'}(\rho)U(\rho),
\end{split}
\end{equation}
where $U$ is either the intralayer or the interlayer interaction, and we have defined $f_{jj'}(\rho) \equiv \rho^{2|m|}e^{-(\beta_{j}+\beta_{j'})\rho}$. Considering the Fourier transforms
\begin{subequations}
\begin{equation}
    f_{jj'}(\rho) = \int \frac{d^{2}q}{(2\pi)^{2}}\, e^{i\qq\cdot\bvec{\rho}}f_{jj'}(q),
\end{equation}
\begin{equation}
    U(\rho) = \int \frac{d^{2}q}{(2\pi)^{2}}\, e^{i\qq\cdot\bvec{\rho}} U(q),
\end{equation}
\end{subequations}
where the expressions for $U(q)$ are given by \eqref{KeldPot}, and
\begin{equation}
\begin{split}
    f_{jj'}(q) =&\, \frac{2\pi}{(\beta_{j} + \beta_{j'})^{2|m|+2}}\Gamma(2|m|+2) {}_{2}F_{1}\bigg( |m|+1,|m| + \frac{3}{2}; 1; -q^{2}(\beta_{j}+\beta_{j'})^{-2}\bigg),
\end{split}
\end{equation}
with ${}_{p}F_{q}(a_{1},\dots,a_{p};b_{1},\dots,b_{q};x)$ the generalized hypergeometric function. Substituting into \eqref{IntKeld}
\begin{equation}
    U_{jj'}^{mm'} = \frac{\delta_{m,m'}}{(2\pi)^{2}}\int_{0}^{\infty}dq\, qf_{jj'}(-q)U(q).
\end{equation}
This integral is well behaved and can be evaluated numerically. Finally, the overlap matrix elements are
\begin{equation}
\begin{split}
    S_{jj'}^{mm'} = \int d^{2}\rho\, \phi_{jm}^{*}(\bvec{\rho})\phi_{j'm'}(\bvec{\rho}) = \delta_{m,m'}\frac{\Gamma(2|m|+2)}{(\beta_{j} + \beta_{j'})^{2|m|+2}}.
\end{split}
\end{equation}

For each quantum number $\Bar{m}$ we define a range for the $m$ values, i.e. $-m_{\text{max}}\leq m \leq m_{\text{max}}$, and for each $m$ we set $N$ basis functions to build the matrix representations in 
\eqref{GenEig}. For sufficiently large values of $m_{\text{max}}$ and $N$, convergence is obtained at the lowest energy levels, achieving a good approximation to the low-energy exciton spectra. For both X and IX excitons, good convergence was achieved in the first 5 energy levels with $m_{\text{max}}=10$ and $N=30$. Considering an hexagonal boron nitride (hBN) encapsulation for the bilayer, the parameters for the electrostatic potentials are $\varepsilon_{\parallel} = 6.9$, $\varepsilon_{\perp} = 3.7$ \cite{geick1966normal,laturia2018dielectric}, and the screening length considered is $r_{*} = 25$ \AA$\,$ (vacuum value) \cite{PhysRevB.100.115203}.

\section{Optical absorption by excitons}

The radiative formation of excitons is driven by the light-matter interaction
\begin{equation}\label{LMHam}
    H_{\text{LM}} = \frac{e \gamma_{p}}{\hbar c} \sum_{\kk,\lambda}\sum_{\bvec{\xi}} \sqrt{\frac{8\pi\hbar c}{\mathcal{V}\xi}} \Big[ c_{c\lambda}^{\dagger}(\kk+\bvec{\xi}_{\parallel})c_{v\lambda}(\kk)a(\bvec{\xi}) + {\rm H.c.}\Big],
\end{equation}
where the operator $a^{\dagger}(\bvec{\xi})$ creates a photon with wave vector $\bvec{\xi}=\bvec{\xi}_{\parallel}+\bvec{\xi}_{\perp}$, which is separated into its in- and out-of-plane components, respectively; $\gamma_{p} = 5.323$ eV$\cdot{\rm \AA}$ is the momentum matrix element for the phosphorene monolayer at the $\Gamma$ point \cite{PhysRevB.100.115203}; and $\mathcal{V}=\mathcal{SL}$, with $\mathcal{S}$ the sample surface area and $\mathcal{L}$ the height of the optical cavity. To calculate the absorption rate (number of photons per unit of time per unit area) due to hybridized excitons, we employ Fermi's golden rule with a Lorentzian line shape:
\begin{equation}
    \Gamma_{i} = \frac{2\pi}{\hbar}\sum_{f} \big| \braoket{f}{H_{\rm LM}}{i}\big|^{2}\,\frac{\beta/\pi}{(E_{f}-E_{i})^{2}+\beta^{2}},
\end{equation}
We set the initial states containing a single photon $\ket{i}=a^{\dagger}(\bvec{\xi})\ket{\Omega}\equiv \ket{\bvec{\xi}}$, the final states as $\ket{{\rm hX}(\QQ)}_{n}$, and the phenomenological broadening $\beta=5$ meV. Furthermore, due to the spatial separation of the charge carriers in the IX excitons, there is a reduction in the probability of dipolar optical transition of these quasiparticles. Then, only the light interaction with the X components of the hX states is considered. The matrix elements are
\begin{equation}
\begin{split}
    {}_{n}\braoket{\text{hX}(\QQ)}{H_{\rm LM}}{\bvec{\xi}} =\, \frac{e\gamma_{p}}{\hbar c} \sqrt{\frac{8\pi\hbar c}{\mathcal{L}\xi}}X^{*}(\bvec{0})\sum_{\mu,\nu}\big[ \mathcal{A}_{\mu\nu}^{n\,*}(\QQ) + \mathcal{B}_{\mu\nu}^{n\,*}(\QQ) \big]\delta_{\QQ+\mu\gG_{1}+\nu\gG_{2},\bvec{\xi}_{\parallel}},
\end{split}
\end{equation}
where momentum conservation is guaranteed by $\delta_{\QQ+\mu\gG_{1}+\nu\gG_{2},\bvec{\xi}_{\parallel}}$. Since the wave vector of the photons that are in resonance with the hX states correspond to the infrared-visible range ($\sim 0.01$ $\text{nm}^{-1}$), and this is negligible compared to the scale of the mBZ ($\sim 1$ $\text{nm}^{-1}$), optical transitions occur for excitons with approximately zero momentum. This means that only components with $\QQ=\bvec{0}$, $\mu=0$ and $\nu=0$ are considered. In this approximation, Fermi's golden rule takes the form
\begin{equation}
    \Gamma(\bvec{\xi}) \approx \frac{16\pi^{2}\gamma_{p}^{2}}{\hbar \mathcal{L}\xi}\frac{e^{2}}{\hbar c}\big| X(\bvec{0})\big|^{2} \sum_{n} \big| \mathcal{A}_{00}^{n}(\bvec{0}) + \mathcal{B}_{00}^{n}(\bvec{0})\big|^{2}\, \frac{\beta/\pi}{[E_{n}(\bvec{0})-\hbar c\xi]^{2}+\beta^{2}}.
\end{equation}
The total number of absorbed photons is obtained by multiplying this expression by the number of photon states in an infinitesimal range of energy $(\epsilon,\epsilon+d\epsilon)$. Since the reciprocal volume element $4\pi\xi^{2}d\xi$ contains $\mathcal{SL}/(2\pi)^{3}$ states, the number of photons becomes $[\mathcal{SL}/(2\pi^{2}\hbar^{3}c^{3})]\epsilon^{2}d\epsilon$. The resulting absorption rate is
\begin{equation}\label{AbsExc}
    A(\epsilon) \approx \frac{8\gamma_{p}^{2}\epsilon \,d\epsilon}{\hbar^{3}c^{2}}\frac{e^{2}}{\hbar c}\big| X(\bvec{0})\big|^{2} \sum_{n} \big| \mathcal{A}_{00}^{n}(\bvec{0}) + \mathcal{B}_{00}^{n}(\bvec{0})\big|^{2}\, \frac{\beta/\pi}{[E_{n}(\bvec{0})-\epsilon]^{2}+\beta^{2}}.
\end{equation}
In an experimental setup, the energy differential $d\epsilon$ is identified as the detector resolution, which is given a typical value of 1 meV.

\section{Twist-angle dependence of the lowest $\gamma$-point exciton}

To study the twist-angle dependence of the moiré localized states, we assume that the confining potential is described by an anisotropic 2D harmonic oscillator:
\begin{equation}
    V(\rr) = \frac{1}{2}M_{x}\omega_{x}^{2}(\theta)x^{2} + \frac{1}{2}M_{y}\omega_{y}^{2}(\theta)y^{2}.
\end{equation}
In the large-supercell approximation, the wells in the potential landscape maintain a constant depth when varying the twist angle, and their lengths along $x$ and $y$ scale like the moiré periodicity, $\propto \theta^{-1}$. Then, the confinement frequency $\omega_{\zeta}$ ($\zeta = x,y$) scales as $\propto \theta$, and we can write the frequency as
\begin{equation}
    \omega_{\zeta}(\theta) = \omega_{\zeta}^{0} + \chi_{\zeta}\theta,
\end{equation}
or equivalently, the zero-point energy of the confined states has the following $\theta$ dependence:
\begin{equation}
    \frac{\hbar \omega_{x}(\theta)}{2} + \frac{\hbar \omega_{y}(\theta)}{2} = \bigg( \frac{\hbar \omega_{x}^{0}}{2} + \frac{\hbar \omega_{y}^{0}}{2} \bigg) + (\sigma_{x} + \sigma_{y})\theta,
\end{equation}
where we have defined $\sigma_{\zeta}\equiv \hbar \chi_{\zeta}/2$.
\end{widetext}
\end{document}